\documentclass[12pt]{article}

\usepackage[utf8]{inputenc}
\usepackage{mathtools}
\usepackage{amsmath}
\usepackage{amsthm}
\usepackage{amssymb}
\usepackage{tikz}
\usepackage{graphicx, nicefrac}
\usepackage{caption}
\usepackage{authblk}
\usepackage{multirow}
\usepackage{lmodern}
\usepackage{enumitem}
\usetikzlibrary{positioning}
\usepackage{amssymb}
\usepackage{adjustbox}

\usepackage{apptools}
\usepackage[authoryear]{natbib}
\usepackage{fullpage}
\AtAppendix{\counterwithin{prop}{section}}

\makeatletter
\renewcommand*{\NAT@spacechar}{~}
\makeatother

\usepackage[utf8]{inputenc}
\usepackage{lineno,hyperref}
\modulolinenumbers[5]
\usepackage{float}
\usepackage{smartdiagram}
\usepackage{geometry}
\usepackage[utf8]{inputenc}
\usepackage{indentfirst}
\usepackage{graphicx}
\usepackage{algorithm}
\usepackage{algorithmic}
\usepackage{amsmath}
\usepackage{amsfonts}
\usepackage{booktabs}
\usepackage{multirow}
\usepackage{caption}
\usepackage{threeparttable}
\usepackage[figuresright]{rotating}
\newtheorem{lemma}{Lemma}[section]
\usepackage{tikz}
\usepackage{multicol}
\usetikzlibrary{intersections, arrows, decorations.pathmorphing, backgrounds, positioning, fit, petri, calc, graphs, fadings, shapes.multipart, shapes.misc, shapes, arrows.meta, shadows, spy}

\usetikzlibrary{arrows.meta}
\usetikzlibrary{bending}
\usetikzlibrary{chains, decorations.pathreplacing, positioning} 
\usetikzlibrary{positioning} 
\usepackage{pgf}
\usetikzlibrary{arrows,automata}
\usepackage{verbatim}
\usepackage[position=top]{subfig}
\usepackage{natbib}

\allowdisplaybreaks % to allow align environment to be split across pages

\newcommand{\myGlobalTransformation}[2]
{
	\pgftransformcm{1}{0}{0.4}{0.5}{\pgfpoint{#1cm}{#2cm}}
}

\newcommand{\rectan}[3]
{
	\begin{scope}
		\myGlobalTransformation{#1}{#2};
		\draw rectangle  (7,4.5);
	\end{scope}
}

\newcommand{\shortrectan}[3]
{
	\begin{scope}
		\myGlobalTransformation{#1}{#2};
		\draw rectangle  (1,4.5);
	\end{scope}
}

\tikzstyle myBG=[line width=3pt,opacity=1.0]

\newcommand{\graphThreeDnodes}[2]
{
	\begin{scope}
		\myGlobalTransformation{#1}{#2};
		\foreach \x in {1,2,3,4,5,6} {
			\foreach \y in {4.5} {
				\node at (\x,\y) [circle,fill=white,inner sep=0pt,minimum size=0pt, pin=above:$t_{\x}$] {};
			}
			\foreach \y in {1,2,3,4} {
				\node at (\x,\y) [circle,fill=black,inner sep=0pt,minimum size=3pt] {};
			}
		}
		\foreach \y in {1,2,3,4}
		\foreach \x in {0}{
			\node at (\x,\y) [circle,fill=white,inner sep=0pt,minimum size=0pt, pin=left:$n_{\y}$] {};
		}
		
		\foreach \y in {1,2,3,4}
		\foreach \x in {8.5}{
			\node at (\x,\y) [circle,fill=white,inner sep=0pt,minimum size=0pt, pin=left:$n_{\y}$] {};
		}
	\end{scope}
}

\newcommand{\networkNode}[2]
{
	\begin{scope}
		\myGlobalTransformation{#1}{#2};
		\foreach \x in {0.5} {
			\foreach \y in {1,2,3,4} {
				\node at (\x,\y) [circle,fill=black,inner sep=0pt,minimum size=3pt] {};
			}
		}
		\foreach \x in {0.5} {
			\foreach \y in {4.5} {
				\node at (\x,\y) [circle,fill=white,inner sep=0pt,minimum size=3pt, pin=above: projection] {};
			}
		}
	\end{scope}
}

\title{An improved decomposition-based heuristic for truck platooning}
\author[1]{Boshuai Zhao}
\author[2]{Roel Leus}
\affil[1]{\footnotesize ORSTAT, KU Leuven, Leuven, Belgium, boshuai.zhao@kuleuven.be}
\affil[2]{\footnotesize ORSTAT, KU Leuven, Leuven, Belgium, roel.leus@kuleuven.be}
\date{}
\begin{document}

		\maketitle

		\noindent
			\textbf{Abstract:} Truck platooning is a promising transportation mode in which several trucks drive together and thus save fuel consumption by suffering less air resistance. In this paper, we consider a truck platooning system for which we jointly optimize the truck routes and schedules from the perspective of a central platform. We improve an existing decomposition-based heuristic by \cite{Luo2020}, which iteratively solves a routing and a scheduling problem, with a cost modification step after each scheduling run. We propose different formulations for the routing and the scheduling problem and embed these into Luo and Larson's framework, and we examine ways to improve their iterative process. In addition, we propose another scheduling  heuristic to deal with large instances. The computational results show that our procedure achieves better performance than the existing one under certain realistic settings.
		
		%\begin{keyword}

	\section{Introduction}	\label{sec:intro}

	Truck platooning is a transport mode in which several trucks drive together at close distances. Due to the aerodynamics, trucks undergo less air resistance and thus save in fuel consumption. Previous research \citep{Bonnet2000} indicates that energy can be saved by at most 20\% for the trailing trucks (so not for the first one; we subsequently refer to these trucks as the ``following'' trucks) of a truck platoon. This transportation mode used to be difficult to achieve because it requires close distances between trucks. With the advances in autonomous driving technology, however, the synergistic operation (e.g. acceleration, braking, and steering) of multiple trucks can be guaranteed, and truck collisions avoided.
	
	Various associations and academic institutes have recently paid attention to truck platooning. 
	In 2015, the Netherlands organization for applied scientific research (TNO) gave a full report about the benefits and risks of this freight mode, involving different supply chain stakeholders \citep{Janssen2015}. Later, the \citet{TheEuropeanAutomobileManufacturersAssociationACEA2017} provided a roadmap describing how to achieve multi-brand platooning before 2025. In this roadmap, there are three stages of autonomy: in the first stage, drivers have to stay alert while driving; in the second stage, drivers in the following trucks of a truck platoon can rest during the trip; and the third stage is fully autonomous driving.
	Moreover, in 2017, the Transport Department of Singapore cooperated with Scania and Toyota Tsusho to design a truck platooning system for their country \citep{Tekst2017}. 
	In 2017, the Volpe center of the U.S. Transportation Department  tested car platooning in Maryland and believed this mode to be beneficial for saving both time and fuel consumption \citep{Tiernan2017}.
	%Recently, \citet{Scherr2019,Scherr2020} have also discussed the platooning potential for city logistics, and \citet{You2020}, \citet{Xue2021}, and \citet{Chen2021} have examined how to combine platooning technology with container drayage.
	
	Creating a truck platoon is often difficult for trucks because they usually follow different routes and delivery schedules. In most cases, before joining a platoon, drivers have to wait in a station, hub, or highway entrance, and even change original routes to meet other trucks, which unavoidably leads to additional delivery costs and waiting time. Therefore, there is not only a trade-off between the benefits and costs of the platooning mode but also an adjustment for the truck's time schedule. 
	
	We present a small problem instance to illustrate the general delivery process and the functioning of the platooning mode. In this instance, three trucks are required to deliver (travel) from their origin to their destination within a given time window; see  Table~\ref{table:case1} for the details. Each truck can choose its own path. If a truck follows another one in a platoon then it can save 10\% fuel costs. For simplicity, we assume that each truck maintains the same constant speed; the corresponding travel time (expressed in minutes) and delivery cost on each arc are proportional, and equated with the arc length; see Fig.~\ref{fig1} for the road network.

	\begin{table}[t]
		\centering
		\footnotesize
		\caption{Truck details for the example instance}
		\begin{tabular}{@{}cccc@{}}
			\toprule
			Truck & origin & destination & time window         \\ \midrule
			$A$     & 1 & 2                    & 2:00 pm to 3:00 pm  \\
			$B$     & 1 & 3                    & 7:00 pm to 8:00 pm  \\
			$C$     & 1 & 6                    & 7:00 pm to 12:00 pm \\ \bottomrule
		\end{tabular}
		\label{table:case1}
	\end{table}
	
	\tikzstyle{vertex}=[circle,fill=black!25,minimum size=15pt,inner sep=0pt]
	\tikzstyle{selected vertex} = [vertex, fill=red!24]
	\tikzstyle{edge} = [draw,thick,-]
	\tikzstyle{weight} = [font=\small]
	\tikzstyle{selected edge} = [draw,line width=5pt,-,red!50]
	\tikzstyle{ignored edge} = [draw,line width=5pt,-,black!20]

	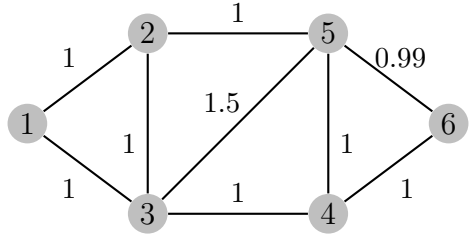
\begin{figure}[t]
		\centering
		
		\begin{tikzpicture}[scale=0.8, auto,swap]
		% Draw a 7,11 network
		% First we draw the vertices
		\foreach \pos/\name in {{(0,1.5)/1}, {(2,3)/2}, {(2,0)/3},
			{(5,0)/4}, {(5,3)/5}, {(7,1.5)/6}}
		\node[vertex] (\name) at \pos {$\name$};	
		% Connect vertices with edges and draw weights
		\foreach \source/ \dest /\weight in {2/1/1,5/2/1, 4/3/1}
		\path[edge] (\source) -- node[weight] {$\weight$} (\dest);

		\foreach \source /\dest /\weight/\pos in {2/3/1/{below left},1/3/1/{below left},4/6/1/{below right},5/6/$\;\;\;0.99$/{above},5/4/1/{below right},3/5/$1.5\;\;\;\;$/{above}} 
		\path[edge] (\source) -- node[weight, \pos] {$\weight$} (\dest);
		
		\end{tikzpicture}
		\caption{Road network for the example instance (the numbers indicate the arc lengths)}\label{fig1}
	\end{figure}

	\begin{figure}[t]
		
		\centering
		%\hspace{0.5cm}
		\begin{minipage}{.45\textwidth}
			
			\begin{tikzpicture}[scale=0.8, auto,node distance = 3cm, % distance between nodes
			thick % line style
			,swap,every node/.style={,fill=yellow!30}]
			
			\foreach \pos/\name in {{(0,1.5)/1}, {(2,3)/2}, {(2,0)/3},
				{(5,0)/4}, {(5,3)/5}, {(7,1.5)/6}}
			\node[vertex] (\name) at \pos {$\name$};	
			
			\foreach \source/ \dest /\weight in { 1/2/ A}
			\draw[arrows=-{>[scale=1, length=4, width=3.5,flex=0]},line width=0.3mm, black ] (\source) -- node[weight] {$\weight$} (\dest);
			
			\foreach \source/ \dest /\weight in {1/2/ A\;C,1/3/B, 2/5/C,5/6/C}
			\draw[arrows=-{>[scale=1, length=4, width=3.5,flex=0]},line width=0.3mm, black ] (\source) -- node[weight] {$\weight$} (\dest);
			\draw[arrows=-{>[scale=1, length=4, width=3.5,flex=0]},line width=0.3mm, black ] (0.15,1.85cm)-- (1.6,2.95cm) ;	
			
			\end{tikzpicture}
			\caption{A feasible solution}\label{fig2}
		\end{minipage}%
		%\hspace{0.5cm}
		\begin{minipage}{.45\textwidth}
			\begin{tikzpicture}[scale=0.8, auto,node distance = 3cm, % distance between nodes
			thick % line style
			,swap,every node/.style={,fill=yellow!30}]
			
			\foreach \pos/\name in {{(0,1.5)/1}, {(2,3)/2}, {(2,0)/3},
				{(5,0)/4}, {(5,3)/5}, {(7,1.5)/6}}
			\node[vertex] (\name) at \pos {$\name$};

			\foreach \source/ \dest /\weight in { 1/2/ A}
			\draw[arrows=-{>[scale=1, length=4, width=3.5,flex=0]},line width=0.3mm, black ] (\source) -- node[weight] {$\weight$} (\dest);
			
			\foreach \source/ \dest /\weight in {3/4/C, 4/6/C}
			\draw[arrows=-{>[scale=1, length=4, width=3.5,flex=0]},line width=0.3mm, black ] (\source) -- node[weight] {$\weight$} (\dest);
			\foreach \source/ \dest /\weight in {1/3/B\;C}
			\draw[arrows=-{>[scale=2, length=4, width=3.5,flex=0]},line width=1mm, black ] (\source) -- node[weight] {$\weight$} (\dest);
			
			\end{tikzpicture}
			\caption{An optimal solution}\label{fig3}
		\end{minipage}
	\end{figure}

	Two solutions are shown to illustrate how platooning influences distribution planning. Fig.~\ref{fig2} depicts a feasible solution in which each truck chooses its shortest path: truck~$A$ follows path~$1-2$, truck~$B$ on path~$1-3$, and truck~$C$ on path~$1-2-5-6$. Since trucks~$A$ and $B$ have non-overlapping delivery intervals, they cannot meet on arc $(1, 2)$ to form a platoon. Therefore, the total delivery cost can be computed as 1 (truck~$A$) + 1 (truck~$B$)  + (1 + 1 + 0.99) (truck~$C$) = 4.99. An optimal solution is given in Fig.~\ref{fig3}, where truck~$A$ still follows path~$1-2$ and $B$ path~$1-3$, but truck~$C$ now takes path~$1-3-4-6$. In this way, trucks~$B$ and $C$ can form a platoon on arc~$(1,3)$, thus saving  $1*10\% = 0.1$ due to the 10\% cost savings for one of the two trucks. The total delivery cost is then as follows: 1 (truck~$A$) +  1 (truck~$B$) + (1 + 1 + 1) (truck~$C$) $- $ $0.1$ (fuel saving) = 4.9.
	
	Our work is inspired by the popular ``sharing economy" concept. The popularity of collaborative shipping is increasing, with as main goal to share the available truck capacity \citep{Creemers2017} and individual logistics capabilities \citep{Dahle2019}. Truck platooning aims to share freight routes to reduce air resistance, and can thus be regarded as a new type of collaborative shipping. The threshold for this type of collaboration is lower because different sources do not need to be in the same truck. Its potential users could be all trucks involved in the freight market, similar to all private cars in the shared travel market. 
	
	In this paper, we focus on the operational planning issues of truck platooning from the perspective of a central platform. Our goal is to explore which benefits can be gained by an integrated truck fleet system, and how to design an algorithm for the system to handle the dispatching of a large number of trucks.
	From a practical perspective, our work can help the freight industry to efficiently integrate freight resources and contributes to the reduction of air pollution. 
	From an academic perspective, this research provides a better method for handling the large-scale joint routing and scheduling problem.
	
	Our contributions are threefold. First, we compare two different mixed-integer programming (MIP) formulations for making truck platooning decisions. Second, we improve a heuristic based on a decomposition framework of \citet{Luo2020}. Specifically, we propose different formulations for the two subproblems of the framework (the routing and the scheduling problem) and improve the iterative process. Third, we provide a heuristic for the scheduling subproblem to handle large instances. 
	
	The remainder of this article is organized as follows: in Section~\ref{liter}, we present a literature review. Subsequently, in Section~\ref{problemandmodel}, we define the truck platooning problem and examine two types of existing formulations, one in continuous time and one in discrete time. Next, a decomposition-based heuristic based on \citeauthor{Luo2020}'s framework is described in Section~\ref{decomposemodel} and a heuristic for large scheduling instances is developed in Section~\ref{pre}. Then, in Section~\ref{numericaldiscussion}  we analyze our computational experiments. Finally, we conclude the article in Section~\ref{conclusion}.
	
	\section{Literature review}	\label{liter}
	
	Below we survey earlier related work on truck platooning.  We first consider the routing problem, then the scheduling problem, and then joint routing and scheduling.
	
	\subsection{Routing problem for truck platooning}
	
	The routing problem for truck platooning is to optimize the vehicle routes to facilitate the formation of truck platoons. %\textcolor{blue}{ Unlike the use of platoons in the vehicle routing problem \citep{Nasri2018} where a truck can lead, the current routing problem means that a set of trucks needs to cooperate with each other on different road sections to save fuel consumption and achieve the lowest cost.}
	\citet{Larsson2015} study the truck platooning problem by ignoring the constraints of delivery deadlines. They prove that their problem is NP-hard and propose two constructive heuristics and a local search algorithm. The first one, called the best pair heuristic, with the same logic as \citet{Larson2013}, chooses the best pairwise platoons based on the platooning savings. The second one, called the hub-based heuristic, divides the trucks into several subsets and selects a hub for each truck subset. Here, the hub is selected from certain nodes traversed by platoons. This method restructures the entire problem into multiple sub-problems, where each sub-problem optimizes the routes for a subset of trucks from the origin to the hub and then from the hub to the destination. 
	
	\subsection{Scheduling problem for truck platooning}
	
	The scheduling problem is to adjust the freight schedule of trucks with fixed routes so as to enable them to form platoons. \citet{Boysen2018} consider the identical-path platooning problem, in which all trucks have the same path but different delivery time windows. They propose three different functions for the  platooning costs (linear, concave, and general) and analyze the computational complexity under different settings. Moreover, they explore the impact of various factors and find that few platooning partners, limited platoon size, and tight delivery time windows can decrease the potential platooning benefits.
	\citet{VandeHoef2016} studies a similar problem but assumes that trucks can drive at different speeds. 
	He first builds a number of truck pairs and then constructs multi-truck platoons based on these pairs. He proves that the problem is NP-hard and proposes a local improvement heuristic for large instances. 
	\citet{Zhang2017} consider platooning with uncertain travel times,  with the objective cost composed of the travel time cost, schedule deviation penalties, and fuel cost. Their results show that platooning is beneficial only when the difference in scheduled arrival times is less than a certain threshold, and uncertainty reduces the threshold.
	\citet{Larsen2019} study the hub-based truck platooning problem where truck platoons are formed at a platooning hub. The chauffeurs are required to take necessary rest after a period of driving (specifically, a rest of 45 min after 4.5 h of driving) and can rest if they are followers in platoons. A dynamic-programming-based local search heuristic is proposed in the study. Results imply that the platooning benefits are relatively limited unless chauffeurs can rest as followers during a trip.
	
	\subsection{Joint routing and scheduling problem for truck platooning}
	
	The joint routing and scheduling problem optimizes the vehicle routes and time schedules together.  
	\citet{Larson2016} conduct research on this joint problem and mainly focus on reducing the problem complexity. They establish a bound for the maximal detour per vehicle, beyond which the additional fuel expenses will be higher than the maximal potential fuel savings. With this bound, they construct a prepossessing procedure to exclude unnecessary paths. For the greater Chicago highway network, their experiment achieves a 1\% optimality gap in less than 300 seconds of runtime. Later, \citet{Luo2018} have extended the model of \citet{Larson2016} to allow different speeds. They propose a MIP model and design a decomposition method, clustering first and routing second, to deal with larger instances. The work by \citet{Luo2020} is the closest to the current paper. 
	The differences with our work are the assumption that the leading truck can also gain fuel savings, and the fact that trucks are not allowed to wait during their trip. 
	\citeauthor{Luo2020} propose an iterative heuristic by solving the routing problem first and then scheduling. They generate valid inequalities to strengthen the formulation and help find high-quality solutions for a large German network consisting of 647 nodes and 1490 arcs. 
	
	Contrary to the above-mentioned studies, which all use continuous-time settings, other authors have  adopted a discrete-time approach and use time-space networks. 
	\citet{Nourmohammadzadeh2019} study the joint problem by proposing a meta-heuristic based on ant colony optimization. Their results indicate that their heuristic is superior to their previous algorithm (a genetic algorithm developed in \citealp{10.1007/978-3-319-49001-4_4}), both in terms of solution quality as well as computational time.
	\citet{Crainic2020} take regulatory break times into account. They formulate a MIP model based on a time-space network and propose a pre-processing procedure to reduce the problem size. They also indicate that the platooning benefits of different automation stages are different. The second stage, where the followers can rest during the journey, can bring the largest benefit. \citet{KishoreBhoopalam2020} mainly focus on 
	the setting where the maximum platoon size is two. They provide a polynomial algorithm for this case and, based on this, they design two fast heuristics for the multi-truck platooning problem. They also perform numerical tests on a Dutch highway network consisting of 20 cities and 45 road sections; the results indicate that two-truck platoons can capture most of the potential platooning savings. \citet{ABDOLMALEKI202191} try to schedule a given set of multi-class trucks with flexible routes and multi-speeds. They model the problem as a concave-cost network problem and propose several solution methodologies, including an outer approximation algorithm (as an exact algorithm), a dynamic-programming-based heuristic, and an approximation algorithm. Numerical results illustrate the efficiency of these algorithms.

	Table~\ref{table:literature} lists several of the closest studies with flexible routes and strict time windows. Our problem statement is the same as in \citet{KishoreBhoopalam2020} and \citet{Nourmohammadzadeh2019}, and is a special case of the problem studied by \citet{ABDOLMALEKI202191}. However, \citet{KishoreBhoopalam2020} mainly consider the case where the platoon size is 2 or 3 and focus on small networks. \citet{ABDOLMALEKI202191} aim to obtain a good solution quickly rather than a near-optimal one. 
	
	Existing studies use either a continuous-time model or a discrete-time model, but, to the best of our knowledge, to date, there has not yet been a comparison of these two types. Moreover, only a few studies focus on solving instances with many trucks and large networks, while this will exactly be our focus. For this setting, we will identify the most suitable formulations and design more effective heuristics.

	\begin{table}[H]
		\centering
		\scriptsize
		\caption{The most closely related papers}
		\setlength{\tabcolsep}{1mm}
		\begin{threeparttable}
			\begin{tabular}{@{}lcccll@{}}
				\toprule
				& Wait\tnote{1}      & Time\tnote{2} & Size\tnote{3}& Network\tnote{4}                                                                                                & Algorithm\tnote{5}                                                                       \\ \midrule
				\citet{KishoreBhoopalam2020} & $\bullet$ & D    & $\bullet$           & \begin{tabular}[c]{@{}c@{}}Dutch (N: 20, A: 45)\end{tabular}                                    & \begin{tabular}[c]{@{}c@{}}Exact: pair platoon; Heur\end{tabular} \\\cmidrule(l){1-6}
				\cite{ABDOLMALEKI202191}            & $\bullet$ & D    &  & \begin{tabular}[c]{@{}c@{}}Random, German (N: 647, A: 1390)\end{tabular} & \begin{tabular}[c]{@{}c@{}}Exact: OA; Heur: DPH;      AP\end{tabular}     \\\cmidrule(l){1-6}
				\begin{tabular}[c]{@{}c@{}}\begin{tabular}[l]{@{}l@{}} Nourmohammadzadeh\\ \& Hartmann (2019)  \end{tabular}\end{tabular}         & $\bullet$ & D    &           & \begin{tabular}[l]{@{}l@{}}Grid, random, and Sweden\\ (N: 356, A: 816) \end{tabular}        & Heur: LSH \\\cmidrule(l){1-6}
				\citet{Luo2020}   &           & C   &  $\bullet$          & \begin{tabular}[c]{@{}c@{}}German (N: 647, A: 1390)\end{tabular}                        & Heur\\\cmidrule(l){1-6}
				\textit{This paper}     & $\bullet$ & D    &   $\bullet$         & \begin{tabular}[c]{@{}c@{}}Grid, German (N: 647, A: 1390)\end{tabular}     & Heur  \\ \bottomrule
			\end{tabular}
			\begin{tablenotes}   
				\scriptsize 
				\item[1]  Wait: the vehicle can wait at the node during the journey
				\item[2] C: continuous time; D: discrete time
				\item[3] Size: platoon size limit
				\item[4] N: node number; A: arc number  
				\item[5] Exact: exact algorithm; Heur: heuristic; DPH: dynamic-programming-based heuristic; LSH: local search heuristic; OA: outer approximation; AP: Approximation algorithm
			\end{tablenotes}
		\end{threeparttable} 
		\label{table:literature}
	\end{table}
	
	\section{Problem statement and linear formulations}\label{problemandmodel}

	In this section, we describe our problem statement (in Section~\ref{sec:problemstat}) and present two MIP formulations for the truck platooning problem (TPP), namely the formulation of \citet{Larson2016} (Section \ref{basicmodel}) and the formulation of \citet{ABDOLMALEKI202191} presented in Section \ref{timespace}. We wish to compare which one performs better and which scenarios they are suitable for from a computational perspective. The computational results will be reported in Section~$\ref{numericaldiscussion}$. 
	
	\subsection{Problem statement} \label{sec:problemstat}
	
	Our goal is to find a high-quality delivery plan for a set of trucks from the perspective of a central platform, and we do not consider the profit distribution over the drivers in a truck platoon. Our model is based on a number of assumptions, with the most important ones as follows. We assume to be working in the first stage of  automation, where each truck needs a driver. All trucks have the same specifications, so the following trucks all enjoy the same cost-saving ratio when driving in a platoon. All trucks also maintain the same constant speed during all trips. Truck platoons can only be formed at nodes in the network (such as stations or rest stops), and cannot be formed while driving. To avoid confusion, if several vehicles form a truck platoon then the vehicle with the smallest index is regarded as the leading truck. There is no limit to the platoon size. In our work, the terms ``leading truck'' and ``following truck" are only used to describe
	the vehicle position in a truck platoon and is completely unrelated to power transmission. 
	
	Our detailed problem statement is as follows: we wish to obtain a distribution plan that minimizes the total delivery cost of all trucks. In graph $G(N, A)$, each truck $v$ has to travel from its origin $O_v$ to destination $D_v$ within its delivery time period $(T^{ed}_v, T^{la}_v)$. If a truck $v$ meets at least one other truck at node $i$, they can form a truck platoon on arc $(i, j)$. The delivery cost on arc $(i, j)$ is $c_{ij}$, but this cost is is reduced by the ratio $\eta$ for all following trucks in a platoon. An overview of the sets and parameters is given in  Table~\ref{table:notation3.0}.

	\begin{table}[h]
		\footnotesize
		\centering
		\caption{Sets and parameters}
		\begin{tabular}{@{}ll@{}}
			\toprule
			Set                                       & Definition                                                                           \\ \midrule
			$N$                                       & Node set\\
			$A$                                       & Arc set\\
			$V$                                       & Vehicle (truck) set\\\midrule
			Parameter                                 & Definition                                                         \\\midrule
			$c_{ij}$                                  & Delivery cost on arc $(i, j){\in}A$\\
			$\eta$                                    & Platooning saving ratio for a following truck\\
			$Q$                                    		& Platoon size limit, maximum vehicle number in a truck platoon\\
			$T_{ij}$                                  & The corresponding  travel time on arc $(i, j){\in}A$\\
			$O_v$                                     & Origin point of vehicle $v{\in}V$, with $O_v{\in}N$\\
			$D_v$                                     & Destination point of vehicle $v{\in}V$, with $D_v{\in}N$\\
			$T^{ed}_v$                                & The earliest departure time for vehicle $v\in V$ at its origin\\
			$T^{la}_v$                                & The latest arrival time for vehicle $v\in V$ at its destination\\
			$st_{ij}$                                & The shortest travel time from node $i\in N$ to node $j\in N$\\
			$\underline{t}_{vi}$, $\overline{t}_{vi}$ & The earliest and latest feasible time for vehicle $v\in V$ to\\
			& \quad enter node $i\in N$; $\underline{t}_{vi}=T^{ed}_v+st_{O_vi}$,   $\overline{t}_{vi}=T^{la}_v-st_{iD_v}$ \\
			\bottomrule
		\end{tabular}
		\label{table:notation3.0}
	\end{table}

	\subsection{Coordinated platooning formulation}\label{basicmodel}
	
	The formulation in this section is from \citet{Larson2016}.  
	Since this formulation aims to coordinate vehicles to form truck platoons, we will refer to it as  \emph{coordinated platooning formulation} (CPF). The decision variables for CPF are given in Table~\ref{table:notation3.1}.
	The formulation can be stated as follows:
	\begin{table}[H]
		\footnotesize
		\centering
		\caption{Decision variables for CPF}
		\begin{tabular}{@{}ll@{}}
			\toprule
			Decision variables 						 &Definition                   \\\midrule
			$x_{ijv}$                               & = 1 if vehicle $v$ traverses arc $(i,j)\in A$; = 0 otherwise; $v\in V$\\
			$y_{ijvw}$                               & = 1 if vehicles $v$ and $w$ form a platoon on arc $(i,j)\in A$ and vehicle $v$ is \\&leading; = 0 otherwise; $v,w\in V$\\
			$t_{iv}$                                 & The time when vehicle $v\in V$ enters node $i\in N$; $T^{ed}_v \le t_{iv} \le T^{la}_v$\\ \bottomrule
		\end{tabular}
		\label{table:notation3.1}
	\end{table}
	
	\begin{footnotesize}
	\label{originalmodel}
	\begin{align}\min \sum_{i,j\in A}\sum_{v\in V} c_{ij} (x_{ijv}-{\eta}\sum_{w\in V}y_{ijvw})\label{TPP0} \end{align}
	\begin{align}
	s.t.&& \sum_{j:(i, j){\in}A}x_{ijv}-\sum_{j:(j, i){\in}A}x_{jiv}
	&= 
	\left\{
	\begin{array}{lll}
	1& i=O_v \\
	-1& i=D_v \\
	0 & otherwise 
	\end{array}\right.												&&v\in V, i\in N\label{TPP1}\\
	&&y_{ijvw}						&\le  		x_{ijw} 				&&v, w{\in}V: v<w, (i, j){\in}A\label{TPP2.1}\\
	&&y_{ijvw}						&\le  		x_{ijv}			 		&&v, w{\in}V: v<w, (i, j){\in}A\label{TPP2.2}\\
	&&(t_{iw}-t_{iv})				&\le  	M^0_{vwi} (1-y_{ijvw}) 	&& v, w{\in}V: v<w, i{\in}N\label{TPP3}\\
	&&(t_{iv}-t_{iw})				&\le  	M^1_{vwi} (1-y_{ijvw})	&& v, w{\in}V: v<w , i{\in}N\label{TPP4}\\
	&&\sum_{v{\in}V: v<w}y_{ijvw}	&\le 		1						&& w{\in}V , (i, j){\in}A\label{TPP5}\\
		&&\sum_{w{\in}V: w>v}y_{ijvw}	&\le  (Q-1)(1-\sum_{u{\in}V: u<v}y_{ijuv})	&&v {\in}V, (i, j){\in}A\label{TPPplatoonsize}\\
	&&t_{jv}-t_{iv}+M^2_{ijv} (1-x_{ijv})	&\ge		T_{ij}					&& v{\in}V , (i, j){\in}A: j\neq O_v, D_v\label{TPP6}\\
	&&T^{ed}_v						&\le 		t_{O_vv}				&& v{\in}V  \label{TPP7}\\
	&&t_{D_vv}				&\le		T^{la}_v 				&&v{\in}V  \label{TPP8}\\
	%&&\sum_{(i, j){\in}A}T_{ij}x_{ijv} &\le		T^{la}_v -T^{ed}_v		&&v{\in}V \label{TPP9}\\
	&&x_{ijv},y_{ijvw}				&\in		\{0, 1\}  				&&v, w{\in}V: v<w , (i, j){\in}A\label{TPP10}\\
	&&t_{iv}						&\ge		0		  				&&v{\in}V\label{TPP11}
	\end{align}
	\end{footnotesize}
	
	The objective function (\ref{TPP0}) minimizes the total delivery cost. The first term in function (\ref{TPP0}) is the fuel consumption of vehicles without platooning benefits; the second term in function (\ref{TPP0}) is the fuel savings due to the platooning mode. 
	Constraints (\ref{TPP1}) ensure that all vehicles travel from their origin to their destination.
	Constraints (\ref{TPP2.1})-(\ref{TPP5}) describe the conditions of platoon formation.
	Constraints (\ref{TPP2.1}) and (\ref{TPP2.2}) imply that a truck platoon consists of at least two vehicles.
	Constraints (\ref{TPP3}) and  (\ref{TPP4}) indicate that the vehicles in a truck platoon should be consistent in time and space, that is, they enter the same arc at the same time. The big-M values $M^0_{vwi}$ and $M^1_{vwi}$ are constants; $M^0_{vwi}=\overline{t}_{iw}-\underline{t}_{iv}$ and $M^1_{vwi}=\overline{t}_{iv}-\underline{t}_{iw}$.
	Constraints (\ref{TPP5}) state that a vehicle participating in a truck platoon has at most one leading truck.
	Constraints (\ref{TPPplatoonsize}) set the maximum platoon size. If vehicle $v$ (the smallest vehicle index) is the leading truck, its following vehicles $w$ would be at most $(Q-1)$. If vehicle $v$ is not a leading truck (but vehicle $u$ is), vehicle $v$ will have no following trucks.
	Constraints (\ref{TPP6}) require that a vehicle must spend at least the appropriate travel time when crossing an arc.
	The big-M values $M^2_{ijv}$ are constants, $M^2_{ijv}=max(0, \overline{t}_{iv}-\underline{t}_{jv}+T_{ij})$.
	Constraints (\ref{TPP7}) and (\ref{TPP8}) set the earliest departure time and the latest arrival time for vehicles. 
	Constraints (\ref{TPP10}) and (\ref{TPP11}) specify the domains of $x$, $y$, and~$t$.

	\subsection{Time-space formulation}\label{timespace}

	In this section, we present the time-space formulation (TSF), modified from \citet{ABDOLMALEKI202191}. The authors formulate their problem as a concave-cost multi-commodity network flow problem that is applicable to a generalized setting, while we formulate our problem similar to a fixed-charge network flow problem, which is a special case of their model. Contrary to \citet{ABDOLMALEKI202191}, however, we do impose limits on the platoon size. Unlike the model CPF which uses continuous time, the TSF is based on a discretization of the time horizon into $T_{max}$ time periods.
	Similar modeling choices have been made in maritime logistics \citep{Zhen2019} and rail transport \citep{Miranda2022}, where the fixed cost can represent a ship or locomotive, while the unit cost can be related to a container or wagon.

	\begin{table}[t]
		\footnotesize
		\centering
		\caption{Sets and parameters for TSF}
		\begin{tabular}{@{}ll@{}}
			\toprule
			Set                        & Definition                                                                                                             \\ \midrule
			$T$                        & The set of time nodes\\
			$N_{ts}$                   & The set of time-space nodes\\
			$A_{all}$                  & The set of time-expanded arcs; $A_{all}=A_{time}\cup A_{ts}$\\
			$A_{time}$                 & The set of time arcs \\
			$A_{ts}$                   & The set of time-space arcs\\
			$\alpha_p^+$               & The set of the time-expanded arcs originating from time-space node $p\in   N_{ts}$\\
			$\alpha_p^-$               & The set of the time-expanded arcs ending at time-space node $p\in   N_{ts}$\\\midrule
			Parameter                  & Definition                                                         \\\midrule
			$tsO_v$                    & Time-space origin of vehicle $v{\in}V$ (with  $tsO_v\in N_{ts}$)            \\
			$tsD_v$                    & Time-space destination of vehicle $v{\in}V$ (with  $tsD_v\in N_{ts}$) \\
			$C_{pq}$                   & Delivery cost on the time-space arc $(p,q) \in A_{ts}$\\
			$C^f_{pq}={\eta} C_{pq}$   & Fixed cost on time-space arc $(p,q)\in A_{ts}$\\
			$C^u_{pq}=(1-\eta) C_{pq}$ & Unit cost on time-space arc $(p,q)\in A_{ts}$\\\bottomrule
		\end{tabular}
		\label{table:notation3.2}
	\end{table}
	
	For this model, we build a time-space network $G(N_{ts}, A_{all})$ based on the physical network $G(N, A)$ and the discretized time horizon.
	Extra sets and parameters are presented in Table~\ref{table:notation3.2}.
	The time-space network $G(N_{ts}, A_{all})$ consists of the time-space node set $N_{ts}$ and the time-expanded arc set $A_{all}$. 
	Each time-space node $p_{it}\in N_{ts}$ indicates a physical node $i\in N$ at time $t\in T$. 
	The time-expanded arc set $A_{all}$ consists of the time arc set $A_{time}$ and the time-space arc set $A_{ts}$. 
	A time arc $a\in A_{time}$ is to connect a time-space node $p_{it}\in N_{ts}$ with another one $q_{i(t+1)}\in N_{ts}$, and passing through it means staying at a physical node $i\in N$ from time $t\in T$ to time $t+1\in T$ without cost. A time-space arc $a\in A_{ts}$ is to connect a time-space node $p_{it}\in N_{ts}$ with $q_{j(t+T_{ij})}\in N_{ts}$; its traversal indicates moving from a physical node $i\in N$ to $j\in N$ in space and from time $t\in T$ to $t+T_{ij}\in T$  with cost $C_{pq}=c_{ij}$. 
	The  time-space origin $tsO_v$ of a vehicle~$v$ represents its physical origin $O_v$ and earliest departure time $T^{ed}_v$; its time-space destination $tsD_v$ represents its physical destination $D_v$ and latest arrival time $T^{la}_v$.

\begin{figure}[t] 
		\centering
		\includegraphics[scale=0.35]{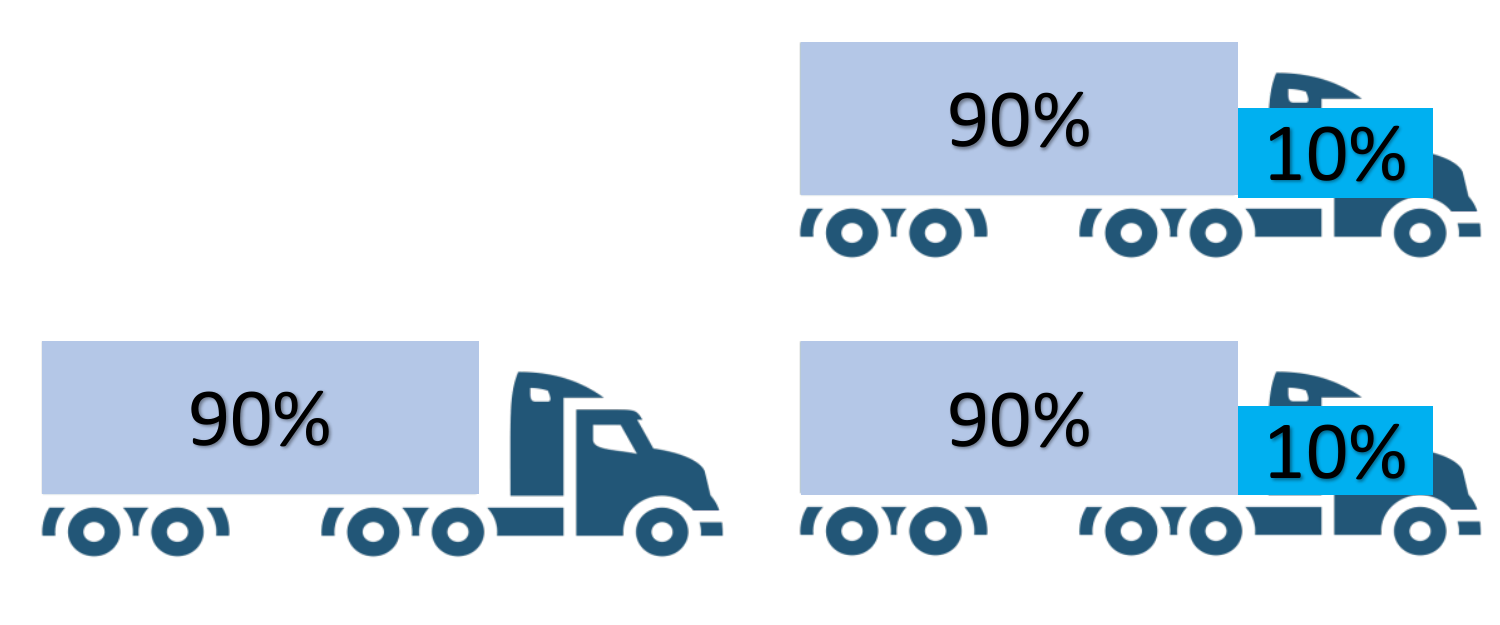}
		\caption{Cost structure for a truck and a truck platoon ($\eta = 10\%$; unit cost: 90\%, fixed cost: 10\%)}
		\label{fig:vp}
	\end{figure}

	The intuition behind TSF is the following. In graph $G(N_{ts}, A_{all})$, each truck $v$ has to travel from its time-space origin $tsO_v$ to its time-space destination $tsD_v$. If a single truck $v$ traverses a time-space arc $a\in A_{ts}$, it incurs a fixed cost $C^f_{a}$ and a unit cost $C^u_{a}$. If $n$ trucks traverse the space-time arc, they can form $\lceil n/Q \rceil$ platoons and pay $\lceil n/Q \rceil$ times the fixed costs. The TPP is thus modeled similarly to the fixed-charge network flow problem. The cost structure is illustrated in Figure~\ref{fig:vp}.
	
	\begin{table}[t]
		\footnotesize
		\centering
		\caption{Decision variables for TSF}
		\begin{tabular}{@{}ll@{}}
			\toprule		
			Decision variables         & Definition          \\\midrule
			$x_{pqv}$                  & = 1 if vehicle $v\in V$ traverses arc $(p,q)\in A_{all}$; = 0 otherwise\\
			$y_{pq}$                   &  $y\in \mathbb{N}$, indicating the platoon number on $(p,q)\in A_{ts}$ \\ \bottomrule
		\end{tabular}
		\label{table:variablesTSF}
	\end{table}
	
	The decision variables for TSF are described in Table~\ref{table:variablesTSF}.
	The relevant range of the indices $p$ and $q$ of the decision variables $x_{pqv}$ can be reduced: for each $p_{it}\in N_{ts}$, the time index $t$ can be restricted to $ \{t\in \mathbb{N}\mid \underline{t}_{iv}\le t \le \overline{t}_{iv}$, $v\in V$, $i\in N\}$.  
	The formulation TSF then looks as follows: 
	\begin{align}
	\min \sum_{(p,q)\in A_{ts}}C^{f}_{pq} y_{pq}+\sum_{(p, q)\in A_{ts}}\sum_{v\in V} C^{u}_{pq} x_{pqv}\label{tsobj}
	\end{align}
	\begin{align}
	s.t.&&\sum_{q\in \alpha_p^+}x_{pqv}-\sum_{q\in \alpha_p^-}x_{qpv}
	&=
	\left\{
	\begin{array}{lll}
	1& p=tsO_v \\
	-1& p=tsD_v \\
	0 & otherwise 
	\end{array}\right.										&&v\in V , p\in N_{ts}\label{tsnetworkflow}\\
&&\sum_{v\in V} x_{pqv}	&\le Q y_{pq}	&&(p, q){\in}A_{ts}\label{tsplatoonsize}\\
	&&x_{pqv}								&{\le}  y_{pq} 		&&v{\in}V , (p, q){\in}A_{ts}\label{tsfixedcost}\\
	&&x_{pqv}								&\in\{0, 1\}		&&v{\in}V , (p,q){\in}A_{all}\label{tsdomain1}\\
	&&y_{pq}								&\in \mathbb{N}		&&(p,q){\in}A_{ts}\label{tsdomain2}
	\end{align}

	The objective function (\ref{tsobj}) minimizes the total cost of the central truck platooning system on the time-space network, including the fixed costs and the variable unit costs. Flow conservation constraints (\ref{tsnetworkflow}) ensure that all vehicles travel from their time-space origin to their destination. Constraints (\ref{tsplatoonsize}) reflect that if there are $n$ vehicles on a time-space arc, at least $\lceil n/Q \rceil$ platoons are required. Constraints (\ref{tsfixedcost}) imply that if any vehicle passes through a time-space arc then the arc incurs the fixed cost at least once (these constraints are not necessary but they tighten the formulation). Constraints (\ref{tsdomain1}) and constraints (\ref{tsdomain2}) specify the domain of $x$ and $y$, respectively.

	\section{Decomposition-based heuristic}\label{decomposemodel}
	
\begin{figure}[t]
		\centering
		\includegraphics[scale=0.3]{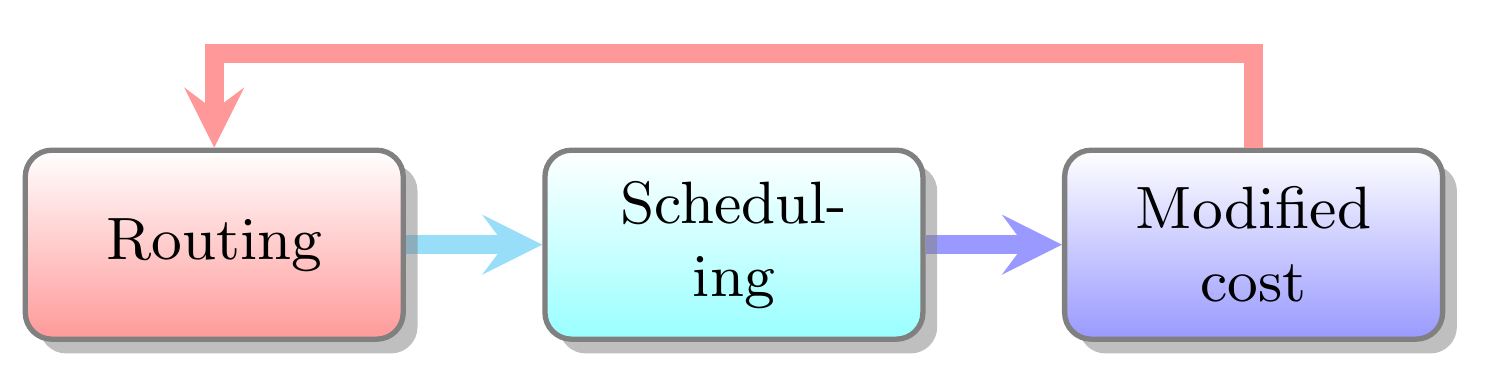}
		\caption{Decomposition logic}\label{decompose}
	\end{figure}

	In order to find high-quality solutions to TPP, we develop a heuristic using the framework of \citet{Luo2020} by iteratively solving routing first and then scheduling (see Fig.~\ref{decompose}). After each scheduling run, the arc costs for the routing stage are updated.
	We present the main differences with  \citet{Luo2020} in  Table~\ref{table:differ}. First of all, \citeauthor{Luo2020} assume that  drivers cannot wait at a node during the journey, which reduces the possibility for trucks to meet other trucks and form platoons, while in our model we do  allow the drivers to wait.  \citet{Luo2020} also argue that the platoon mode benefits both the following and leading vehicles, whereas we, in line with most other studies, assume that the mode only brings advantages to the following vehicles and ignores the relatively low benefit to the leading vehicles.
	\citeauthor{Luo2020} also work in continuous time, while we have a discrete-time model. Methodology-wise, we propose different formulations for the two decomposed problems and we slightly improve the iterative process. For the routing problem, our formulation has fewer variables and constraints, and for the scheduling problem we use a time-indexed formulation while \citeauthor{Luo2020}'s is assignment-based. 
	Finally, we also make a number of changes to the arc cost generation strategy.   
	
	\begin{table}[t]
		\footnotesize
		\centering
		\caption{Differences between \citet{Luo2020} and our work}
		\begin{threeparttable}
			\begin{tabular}{@{}ccccccc@{}}
				\toprule
				\multirow{2}{*}{}   & \multicolumn{3}{c}{Problem}                                                                           & \multicolumn{3}{c}{Methodology (Heuristic)}                                                                                                                                                                                      \\ \cmidrule(l){2-7} 
				&  Wait\tnote{1}                                                                   &Benefit\tnote{2}  & Time\tnote{3} & Routing                                                                   & Scheduling                                                             & Modified cost                                                          \\ \midrule
				\begin{tabular}[c]{@{}c@{}}Luo \& Larson\\(2022)  \end{tabular}& No\ & \begin{tabular}[c]{@{}c@{}}Leader and\\      follower\end{tabular}   & C    & \begin{tabular}[c]{@{}c@{}}Three sets \\of variables\end{tabular} & \begin{tabular}[c]{@{}c@{}}Assignment-based\\ formulation\end{tabular} & \begin{tabular}[c]{@{}c@{}}Three types of\\ cost modifications\end{tabular}         \\\midrule
				\textit{This paper}                       & Yes                                                                & Follower only  & D    & \begin{tabular}[c]{@{}c@{}}Two sets \\of variables\end{tabular}& \begin{tabular}[c]{@{}c@{}}Time-indexed\\ formulation\end{tabular}     & \begin{tabular}[c]{@{}c@{}}Four types of \\cost modifications\end{tabular} \\ \bottomrule
			\end{tabular}
			\begin{tablenotes}   
				\scriptsize              
				\item[1] Wait: the vehicle can wait at a node during the journey
				\item[2] Benefit: the beneficiaries in the truck platoon; Leader: leading truck; Follower: following truck
				\item[3] C: continuous time; D: discrete time
				%\item[4] RS: routing first and then scheduling; ACM: arc cost modification 
			\end{tablenotes}
		\end{threeparttable}
		\label{table:differ}
	\end{table}
	
	In terms of the TSF, we focus on the space dimension to decide the vehicle routes first and then consider the time dimension to optimize the truck schedule. Sections \ref{routeP} and \ref{scheduleP} contain the model for  the routing problem and the scheduling problem, respectively, and the iteration process is discussed in Section \ref{iterationP}.

	\subsection{Routing problem}\label{routeP}

	\begin{figure}[t]
		\begin{tikzpicture}[scale=1]
		
		\rectan{0}{0}{black!50};
		\shortrectan{8.5}{0}{black!50};
		
		\graphThreeDnodes{0}{0};
		\networkNode{8.5}{0};
		%draws special rectangular
		\begin{scope}
		\pgftransformreset
		%time space
		\draw[black, very thick,dashed][arrows=-{>[scale=1, length=4, width=3.5,flex=0]},line width=0.3mm ] (1.4,0.5) -- ++(1,0);
		\draw[black, very thick,dashed][arrows=-{>[scale=1, length=4, width=3.5,flex=0]},line width=0.3mm ] (2.4,0.5) -- ++(1,0);
		\draw[black, thick][arrows=-{>[scale=1, length=4, width=3.5,flex=0]},line width=0.3mm ] (3.4,0.5) -- ++(1.4,0.5);
		\draw[black, very thick,dashed][arrows=-{>[scale=1, length=4, width=3.5,flex=0]},line width=0.3mm ] (4.8,1) -- ++(1,0);
		\draw[black, thick][arrows=-{>[scale=1, length=4, width=3.5,flex=0]},line width=0.3mm ] (5.8,1) -- ++(1.4,0.5);
		
		\draw[blue, very thick,dashed][arrows=-{>[scale=1, length=4, width=3.5,flex=0]},line width=0.3mm ] (1.8,1) -- ++(1,0);
		\draw[blue, very thick,dashed][arrows=-{>[scale=1, length=4, width=3.5,flex=0]},line width=0.3mm ] (2.8,1) -- ++(1,0);
		%\draw[blue, very thick,dashed,->] (3.8,1.1) -- ++(1,0);
		\draw[blue, thick][arrows=-{>[scale=1, length=4, width=3.5,flex=0]},line width=0.3mm ] (3.8,1) -- ++(1.4,0.5);
		
		\draw[green, thick][arrows=-{>[scale=1, length=4, width=3.5,flex=0]},line width=0.3mm ] (3.2,1.5) -- ++(1.4,0.5);
		
		% space
		\draw[black, thick][arrows=-{>[scale=1, length=4, width=3.5,flex=0]},line width=0.3mm ] (9.4,0.5) -- ++(0.4,0.5);
		\draw[black, very thick][arrows=-{>[scale=1, length=4, width=3.5,flex=0]},line width=0.3mm ] (9.8,1) -- ++(0.4,0.5);
		\draw[blue, very thick][arrows=-{>[scale=1, length=4, width=3.5,flex=0]},line width=0.3mm ] (9.7,1) -- ++(0.4,0.5);
		\draw[green, thick][arrows=-{>[scale=1, length=4, width=3.5,flex=0]},line width=0.3mm ] (10.2,1.5) -- ++(0.4,0.5);
		
		%schedule
		%\draw[green, thick,->] (1.9,-0.5) -- ++(1,0);
		\end{scope}
		
		\end{tikzpicture}
		
		\caption{Time-space network and the projection onto the space network}\label{timespaceAndspaceproject}
	\end{figure}
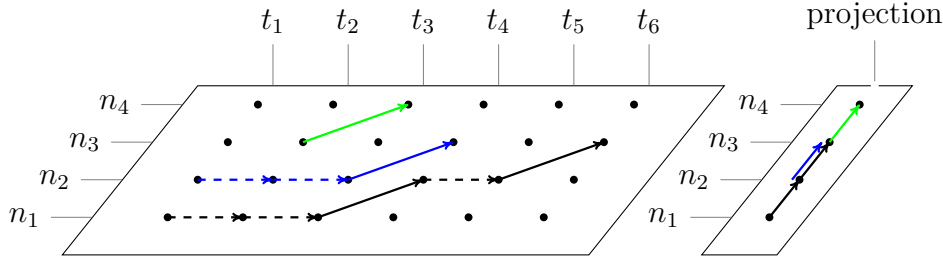
	
	The routing problem is modeled as a fixed-charge network flow problem with time windows. In this routing stage, as long as two trucks drive on the same arc, regardless of whether they meet at the same time, they only need to pay the fixed cost of the arc once (we do not consider platoon size limits here, so fixed cost is incurred at most once). The left part of Fig.~\ref{timespaceAndspaceproject} is a time-space network, in which the total weighted value of these traversed time-space arcs is the target of TPP\@. The right part of Fig.~\ref{timespaceAndspaceproject} is a physical network, onto which the time-space network is projected, and in which the minimum total weighted value of the traversed physical arcs is the target of the routing problem. Therefore, the routing problem can provide a lower bound to TPP.
	
	\begin{table}[t]
		\footnotesize
		\centering
		\caption{Parameters and decision variables for the linear formulation for the routing problem}
		\begin{tabular}{@{}ll@{}}
			\toprule
			Parameter                                 & Definition                                                         \\\midrule
			$c^f_{ij}$                                & fixed cost on arc $(i,j)\in A$, $c^f_{ij}={\eta} c_{ij}$\\
			$c^u_{ij}$                                & unit cost on arc $(i,j)\in A$, $c^u_{ij}=(1-{\eta})c_{ij}$\\\midrule
			Decision variables 		&Definition                   \\\midrule
			$X_{ijv}$                                 & =1 if vehicle $v\in V$ traverses arc $(i, j)\in A$; = 0 otherwise\\
			$Y_{ij}$                                 & =1 if there is at least one vehicle on arc $(i, j)\in A$; =0 otherwise\\
			\bottomrule
		\end{tabular}
		\label{table:notation4.1}
	\end{table}
	
	The decision variables together with some extra parameters for this fixed-charge network flow formulation (FCNF) are provided in Table~\ref{table:notation4.1}.
	The objective function and constraints \label{wholeRP} are as follows: 
	
	\begin{align}
	\min&&\sum_{(i,j)\in A}  c^f_{ij} Y_{ij}+\sum_{v\in V} \sum_{(i,j)\in A} c^u_{ij} X_{ijv}\label{RP0}\\
	s.t.&&\sum_{j:(i, j)\in A}^{n}X_{ijv}-\sum_{j:(j, i)\in A}^{n}X_{jiv}
	&=
	\left\{
	\begin{array}{lll}
	1& i=O_v \\
	-1& i=D_v \\
	0 & otherwise 
	\end{array}\right.	&& i\in N, v\in V\label{RP1}\\
	&& X_{ijv}					&\le  Y_{ij}				&&(i, j)\in A, v{\in}V  \label{RP2}\\
	&&\sum_{(i, j)\in A}T_{ij}X_{ijv} &\le		T^{la}_v -T^{ed}_v		&&v{\in}V \label{RP3}\\
	&& 	X_{ijv}					&\in\{0, 1\}  			&& (i, j)\in A,v{\in}V  \label{RP4}\\
	&& 	Y_{ij}					&\in \{0, 1\}  			&& (i, j)\in A\label{RP5}
	\end{align}

	The objective function (\ref{RP0}) minimizes the total cost of the central truck platooning system, including the fixed cost and the unit flexible cost on all arcs. Flow conservation constraints (\ref{RP1}) ensure that all vehicles travel from their origins to destinations. Constraints (\ref{RP2}) reflect that if any vehicle passes through an arc, the fixed cost on this arc is incurred. Constraints (\ref{RP3}) require that the vehicle paths be compatible with each vehicle's time window. Specifically, the traveling time of each vehicle from its origin to destination should not exceed the length of its time window. Constraints (\ref{RP4}) and constraints (\ref{RP5}) specify that $X$ and $Y$ are binary, respectively.

	Compared with \citet{Luo2020}'s routing formulation (LRF), FCNF has similar network flow constraints for the routing stage  but uses only two sets of decision variables ($X$ and $Y$). LRF uses three sets of variables: the extra ones count the number of additional vehicles on an arc when more than one vehicle traverses this arc, and this also entails extra constraints. We conjecture that fewer variables and constraints can speed up the computations; we verify this computationally in Section \ref{numericaldiscussion}. 
	Additionally, LRF also uses a set of binary variables to know whether at least two vehicles cross the same arc; these variables are unrelated to our problem setting, so they are also removed.

	%$Dist_{i,j}$: the shortest path length from node $i\in N$ to node $j\in N$.
	
	%$E_v$: the set of feasible arcs for vehicle $v\in V$.
	
	%\begin{equation}\label{feasible}
	%E_v=\left\{(i,j)\in A\mid Dist_{O_v, i}+Dist_{j,D_v}\le Dist_{O_v,D_v}/(1-\eta)\right\}
	%\end{equation}

	\subsection{Scheduling problem}\label{scheduleP}
	
	The scheduling problem is to minimize the total fixed cost by adjusting the vehicle schedule on the fixed routes obtained from the routing problem. In Fig.~\ref{timespaceAndspaceproject}, the scheduling problem is to move the solid lines in the left part along the timeline (horizontally) to minimize the total weighted value of the traversed space-time arcs. In this way, after solving the scheduling problem, we can obtain a feasible solution and an upper bound for TPP.

	\begin{table}[t]
		\footnotesize
		\centering
		\caption{Parameters, sets, and decision variables for the scheduling problem}
		\begin{tabular}{@{}lll@{}}
			\toprule
			Parameter                & Definition                                                                                                               \\ \midrule
			$\bar{X}_{ijv}$          & The values  $X_{ijv}$ for an optimal solution to the routing problem   \\%\midrule
			${pt}^v_{ij}$             & The shortest travel time along path $P_v$ from node $i$ to node $j$, $v\in V$, \\&$(i,j)\in P_v$; ${pt}^v_{ij}\ge {st}_{ij}$\\
			$\underline{t}_{vi}$, $\overline{t}_{vi}$         & The earliest and latest entry time for vehicle $v\in V$ on node $i\in P_v$; \\&Here (contrary to Table~\ref{table:notation3.1}
			in Section~\ref{basicmodel}), $\underline{t}_{vi}=T^{ea}_v+pt^v_{O_v i}$,   $\overline{t}_{vi}=T^{la}_v-pt^v_{i D_v}$ \\
			\midrule
			Set                      & Definition \\\midrule
			$P_{v}$                  & The path (set of arcs) of vehicle $v\in V$ obtained from the routing problem; \\
			&  $P_v=\left\{(i,j)\in {A} :   \bar{X}_{ijv}=1\right\}$ %, P_v \subseteq A$                          
			\\
			$P$                      & The union of the arc sets traversed by all vehicles; $P=\cup_{v\in V} P_v$                                    \\%\midrule
			${PT}^{v}_{i}$           & The feasible entry time set for vehicle $v\in V$ on node $i:(i,j)\in P_v$; \\&  
			%and vehicle $v$ can only move along path $P_v$;
			${PT}^{v}_{i}=\left\{t\in \mathbb{N}\mid \underline{t}_{vi} \le t \le   \overline{t}_{vi} \right\}$  \\\midrule
			Decision variable      			 & Definition  \\\midrule
			$y_{ijt}$                & $y\in \mathbb{N}$, indicating the pieces of incurring fixed cost on $(i, j)\in P$ at time $t\in T$\\
			$x^v_{ijt}$              & = 1 if vehicle $v\in V$ enters arc $(i, j)\in P_v$ at time $t\in   {PT}^{v}_{i}$; = 0 otherwise\\ \bottomrule
		\end{tabular}\label{table:notation4.3}
	\end{table}

	The definitions of the  parameters, sets, and decision variables are summarized in Table~\ref{table:notation4.3}. Here, ${pt}^v_{ij}$ describes the shortest time from node $i\in N$ to node $j\in N$ on the fixed path~$P_v$, while previously ${st}_{ij}$ represented the shortest time from node $i\in N$ to node $j\in N$ in the entire graph.
	Our full formulation for the scheduling problem \label{wholeSP} is then as follows: 
	\begin{equation}
	\max\sum_{v\in V} \sum_{(i,j)\in A} c^f_{ij} \bar{X}_{ijv} -\sum_{t\in T}\sum_{(i, j)\in P}c^f_{ij} y_{ijt}\label{SP6}
	\end{equation}
	\begin{align}						
	s.t.
	&&\sum_{\tau=\underline{t}_{vi}}^t {x^v_{ij\tau}}	 	&\ge  \sum_{\tau=\underline{t}_{vj}}^{t+T_{ij}} {x^v_{jk\tau}}
	&& v{\in}V,(i,j), (j,k){\in}P_v, t{\in}T	\label{SP1}\\
	&& \sum_{t\in T} {x^v_{ijt}}					& = 	 1  		&& v{\in}V , (i,j){\in}P_v					\label{SP2}\\
	&& 	\sum_{v\in V}{x^v_{ijt}}										& \le 	Qy_{ijt} 	&& v{\in}V , (i,j){\in}P,t{\in}T	\label{SPlimit}\\
	&& 	{x^v_{ijt}}										& \le 	y_{ijt} 	&& v{\in}V , (i,j){\in}P_v,t{\in}T	\label{SP3}\\
	&& 	{x^v_{ijt}}										&\in 	\{0, 1\}  	&& v{\in}V, (i,j){\in}P_v, t{\in}T		\label{SP4}\\
	&& 	y_{ijt}											&\in \mathbb{N}  	&& v{\in}V , (i,j){\in}P_v 						\label{SP5}
	\end{align}

	The objective function (\ref{SP6}) contains the total platooning savings. The first term is a parameter indicating the total fixed cost of all vehicles without platooning mode (e.g a three-truck platoon leads to three times the fixed cost), while the second term represents the total fixed cost of all vehicles when platooning is possible (e.g a three-truck platoon gives to one  times the fixed cost).
	Constraints~(\ref{SP1}) are precedence constraints between a vehicle's entry time on an arc and the entry time on the successor of this arc.  These are so-called disaggregated precedence constraints for time-indexed formulations \citep{Christofides1987}. We have also tested the aggregated variant of the constraints, but the performance was not as good.
	Constraints (\ref{SP2}) state that each vehicle  enters each physical arc on its path at exactly one time instant.
	Constraints (\ref{SPlimit}) reflect that if there are $n$ vehicles on an arc at a time, at most $\lceil n/Q \rceil$ times of fixed cost.
	Constraints (\ref{SP3}) indicate whether fixed cost on an arc at a given time is generated (these constraints tighten the formulation).
	Constraints~(\ref{SP4}) and (\ref{SP5}) specify the domain of $x$ and $y$, respectively.

	%\textcolor{green}{If a vehicle traverses an arc where no other vehicles pass, or the vehicle cannot meet other vehicles on the arc, it cannot participate in a truck platoon on the arc. Only when a vehicle can find another one that has a coincident time with it on an arc, it has a chance to join a truck platoon. Only the latter case needs to be included in the optimization.}
	
	\citet{Luo2020} apply an assignment-based formulation (ABF) to solve the scheduling problem, while ours is a time-indexed formulation (TIF). The TIF aims to find the minimum usage of the traversed time-space arcs, while ABF tries to match (two or more) vehicles together to maximize cost savings. In another paper, \citet{Luo2022} proposed an adaptive time discretization method for the scheduling problem of \citet{Luo2020}, which dealt with fixed-route truck platooning without allowing drivers to wait during the trip. However, the new method is not suitable for our settings where drivers are allowed to wait during the trip and its formulation aims to maximize cost savings, rather than to minimize the use of arcs (in our approach).
	
	Before optimization of the scheduling stage, we first apply a pre-processing run to  reduce scheduling complexity by selecting vehicles and arcs, based on the following observation:
	for a given vehicle~$v$, if there is another vehicle~$u$ on arc~$(i,j)$ with overlapping time windows on the arc (${PT}^{v}_{i}\cap {PT}^{u}_{i}\neq\emptyset$) then vehicle~$v$ can potentially join a truck platoon.
	Otherwise, vehicle~$v$ must travel alone on arc~$(i,j)$.
	Hence, only the former case needs to be included in the optimization. 
	
	\subsection{Iterative process}\label{iterationP}
	
	The iterative process from \citet{Luo2020} includes three steps (see also Fig.~\ref{decompose}):
	\begin{description}
		\item[Step 1] Solve the routing problem and record vehicle paths.
		
		\item[Step 2] Solve the truck scheduling problem while respecting the vehicle paths from Step 1.
		
		\item[Step 3] Modify the arc costs for the next routing run.
		
	\end{description}
	\noindent Next, return to Step~1 unless any of the following two termination criteria is met: (i) any path from the routing stage is repeated too many times, or (ii) the time limit is reached.
	
	In this section we focus on Step~3 by modifying the arc cost update procedure. In each iteration the objective coefficients (arc costs) of the routing problem for the next stage are updated.  The values are chosen so that each vehicle's arc costs will jointly reflect the actual routing cost as well the timing constraints. 
	We provide an example to better illustrate the intuition behind these cost updates. Consider again the instance presented in Section~\ref{sec:intro} and Fig.~\ref{fig1}, and imagine that there are 10 trucks $A, B, C, \ldots$ that can potentially traverse arc $(1,3)$. In the routing problem in the first iteration, the unit cost and the fixed cost for all vehicles on arc $(1,3)$ are $0.9$ and $\eta = 0.1$, respectively. The routing solution might then suggest that five trucks traverse arc $(1,3)$, and thus each truck gives rise to an average cost of $0.92 =(5*0.9 + 0.1)/5$. Suppose now that in the subsequent scheduling solution, due to timing constraints, only truck~$B$ and truck $C$ form a platoon on arc $(1,3)$, and thus incur an average arc cost of $0.95$ each, namely $(2*0.9 + 0.1)/2)$. In the second-iteration routing problem, the cost of trucks~$B$ and $C$ will then be fixed to $0.95$ on arc $(1,3)$ and there will not be  a combination of unit costs and fixed costs for the next iteration.  In this way, we try to avoid potentially  impossible routing solutions with arc costs of $0.92$.

	\subsubsection{Modifications to the routing objective function}\label{notation-n}
	During the iterative procedure, the scheduling stage remains unchanged (as in Section~\ref{wholeSP}), but we make some minor modifications to the routing problem (compared to Section~\ref{wholeRP}) to incorporate the cost updates. The objective function in the routing problem in iteration~$(n+1)$ is changed as follows:
	
	\begin{equation} \label{RPn0}
	\begin{aligned}
	\min
	\sum_{(i,j)\in A\backslash  P^{(n)}} c^f_{ij} Y_{ij} 
	+\sum_{(i,j)\in A\backslash P^{(n)}}\sum_{v\in V^{(n)}_{ij}} c^u_{ij} X_{ijv}
	+\sum_{(i,j)\in  P^{(n)}}\sum_{v\in V^{(n)}_{ij}} C^{(n+1)}_{ijv} X_{ijv} 
	\end{aligned}
	\end{equation}
	
	Here, $C^{(n)}_{ijv}$, $P^{(n)}$, and $V^{(n)}_{i j}$ are the parameters in iteration~$n$, indicating the modified cost for vehicle~$v$ on arc $(i, j)$, the  set of all arcs traversed by at least one vehicle, and the set of vehicles traversing arc $(i, j)$, respectively. 
	
	The objective function (delivery cost) in iteration~$(n+1)$ is created based on the solution from iteration~$n$ and consists of two parts. The first and the second term represent the cost on arcs not traversed by any vehicle in iteration~$n$, which are the fixed cost and the unit flexible cost, respectively. The third term indicates the cost on the arcs traversed by vehicles in iteration~$n$. On these arcs, each vehicle has its own single arc cost parameter instead of a unit and a fixed cost. 
	
	\subsubsection{Modified cost}

	In this subsection we look into the cost modifications, so the values $C^{(n+1)}_{ijv}$. We will explain the difference between the cost modification procedure of \citet{Luo2020} (LLCMP) and our improved cost modification procedure (ICMP).
	We distinguish four scenarios for the modification, which are summarized in Table~\ref{table:costmodification}. 
	
	\begin{table}[h]
		\footnotesize 
		\setlength{\tabcolsep}{2mm}
		\centering
		\caption{Possible scenarios for modified arc costs for vehicle~$v$ at the end of iteration~$n$}
		\begin{threeparttable}
			\begin{tabular}{@{}clll@{}}
				\toprule
				\multicolumn{1}{l}{Scenario} & Condition for the arc                                                                                      & LLCMP                      & ICMP                                                            \\ \midrule
				1  & traversed by vehicle $v$                                                                                              & real cost    & real cost                                                   \\\cmidrule(l){1-4}
				2  & \multirow{3}{*}{\begin{tabular}[c]{@{}l@{}}traversed by vehicle~$u$ (not $v$), \\ without risk of iterative cycle\end{tabular}} & \multirow{2}{*}{unit cost} & \begin{tabular}[c]{@{}l@{}} if vehicles $u$ and $v$ cannot meet,\\ unit cost + fixed cost \end{tabular}  \\\cmidrule(l){1-1}\cmidrule(l){4-4}
				3  &                                                                                                                       &                            & \begin{tabular}[c]{@{}l@{}}if vehicle $u$ and $v$ can meet,\\ unit cost + partial fixed cost\end{tabular}    \\\cmidrule(l){1-4}
				4  & \begin{tabular}[c]{@{}l@{}}traversed by vehicle $u$ (not $v$),\\with risk of iterative cycle\end{tabular}                  & earlier cost  & earlier cost                                                  \\ \bottomrule
			\end{tabular}
		\end{threeparttable}
		\label{table:costmodification}
	\end{table}
	
	In scenarios~1 and 4, LLCMP and ICMP are the same.
	The first scenario pertains to the arc cost for a vehicle that traverses an arc; the next-iteration arc cost then equals its ``real cost''. If the vehicle joins a platoon, the real cost is the platoon's average cost; otherwise, if the vehicle drives alone, the real cost is the sum of the vehicle's unit cost and fixed cost.
	We follow a slightly different approach in the second and third scenarios. In LLCMP, if vehicle~$u\in V \backslash \{v\}$ passes through an arc~$(i,j)$ but vehicle $v$ does not, then vehicle~$v$'s cost for $(i,j)$ in the next iteration would be the arc's unit cost, since vehicle~$v$ could become the following truck and thus the fixed cost would not be incurred.
	By contrast, we subdivide this situation into two distinct scenarios, depending on whether there is a vehicle~$u\in V$ on arc~$(i,j)$ that can actually meet vehicle $v$ on the arc, meaning that  their time windows on this arc overlap.
	Therefore, in ICMP's scenario~$2$, if there is no other vehicle that can meet vehicle~$v$ on arc~$(i,j)$, the next-iteration arc cost will be the unit cost plus the fixed cost. 	
	In ICMP's scenario~$3$, if there is such a vehicle~$u\in V \backslash \{v\}$ on the arc, then vehicle~$v$ can be a following truck in the next iteration. In that case, we let the next-iteration arc cost of vehicle~$v$ include not only the unit cost but also part of the fixed cost, since the vehicles in a platoon need to share the fixed cost. In this way, the arc cost is normally close to its next-iteration actual cost, which avoids attracting extra vehicles to the arc due to an under-estimation of the costs.
	The fourth scenario is included because  \citet{Luo2020} make the following observation: if a truck platoon in iteration~$k$ made vehicle~$v$ use arc~$(i,j)$ in iteration~$(k+1)$, and a  platoon with the same vehicle composition on this arc formed in iteration~$n$, then it might play the same role again for vehicle $v$ in iteration~$(n+1)$ if the same cost update takes place, and the iterative process might get stuck in a loop. To avoid this, vehicle~$v$ will copy its  cost for arc $(i,j)$ in iteration~$(n+1)$ from iteration~$(k+2)$; this is referred to as ``earlier cost'' in the table.

	\section{A heuristic for large scheduling instances}\label{pre}
	
	\textit{Pairwise} truck platooning is commonly used in truck platooning problems \citep{VandeHoef2016,Liang2016}; this refers to the special case in  which the maximal platoon size is two and each vehicle joins at most one platoon.  We propose a ``pairwise pre-processing heuristic'' to schedule trucks with fixed routes based on pairwise platooning; this heuristic is only used for large instances.  Some extra definitions are provided in Table~\ref{table:notation5.2}.  The details are as follows:

	\begin{table}[h]
		\footnotesize
		\centering
		\caption{Sets, parameters, and decision variables for the scheduling heuristic of Section~\ref{pre}}
		\begin{tabular}{@{}ll@{}}
			\toprule
			Set                                 & Definition                                                         \\\midrule
			$W$                                & the set of potential pairwise platoons $\{u,v\} \subset V $\\
			%			$\delta(v)$                                & the set of vehicles $u\in V$ that can form a  platoon pair with vehicle $v \in V$, so $\{u,v\} \in W$  \\
			\midrule
			Parameter                                 & Definition                                                         \\\midrule
			$s_{\{u,v\}}$                              &  the savings of the pairwise platoon $\{u,v\}\in W$\\
			$\gamma$                                & the maximum number of pairwise platoons, as a fraction of the \\&vehicle number; $0\le \gamma \le 0.5$\\\midrule
			Decision variables 		&Definition                   \\\midrule
			$w_{\{u,v\}}$                                 & $=1$ if the pairwise platoon $\{u,v\}\in W$ is chosen; $=0$, otherwise\\
			\bottomrule
		\end{tabular}
		\label{table:notation5.2}
	\end{table}

	\begin{description}
		
		\item [Step 1: List of potential pairwise platoons and their savings] If two vehicles have a common road segment consisting of several consecutive arcs and overlapping time windows on the segment then they can form a platoon; the pair is then added to set $W$, and their savings are stored as values $s_{\{u,v\}}$. 
		
		\item [Step 2: Select pairwise platoons]  The selection is performed by means of a MIP model that maximizes the total platooning savings. Each vehicle can join only one platoon.  The following formulation is used; this is simply a maximum-weight matching problem with a cardinality constraint.

		\begin{align}
		\max&&\sum_{\{u, v\}{\in}W}  s_{\{u,v\}} w_{\{u,v\}} \label{MM0}\\
		s.t.
		&&\sum_{\{u,v\} {\in} W}	w_{\{u,v\}}	&\le 1  					&& v\in V	\label{MM1}\\
		&&\sum_{\{u, v\}{\in}W}w_{\{u,v\}}		&\le \gamma |V|  				&& 	\label{MM2}\\
		&& 	w_{\{u,v\}}						&\in \{0, 1\}  			&& \{u, v\} \in W	\label{MM3}
		\end{align}
		
		The objective function (\ref{MM0}) is to maximize the total truck platooning savings. 
		Constraints~(\ref{MM1}) state that a vehicle can take part in at most one platoon. 
		Constraints (\ref{MM2}) upper bound the number of selected platoons by $\gamma|V|$. 
		Constraints (\ref{MM3}) specify that variables $w$ are binary.
		
		\item [Step 3: Shorten the time windows for the selected pairs] For each truck in each of the  selected pairwise platoons $\{u,v\}$ from the above model, we change the time window based on the allowed ranges at the merge node $i$ (which is the starting point of the shared road segment). The new time windows for vehicles $u$ and $v$ are ($T^{ed}_u+max(0,\underline{t}_{vi}-\underline{t}_{ui})$, $T^{la}_u-max(0,\overline{t}_{ui}-\overline{t}_{vi})$) and ($T^{ed}_v+max(0,\underline{t}_{ui}-\underline{t}_{vi})$, $T^{la}_v-max(0,\overline{t}_{vi}-\overline{t}_{ui})$), respectively. 
		
		\item [Step 4: solve TIF with shortened time window and unlimited platoon size] During computation, we found that relaxing platoon size constraints might speed up computing and the relaxed problem's optimal solution can serve as a good feasible one for the original TIF. Therefore, we create a modified TIF by relaxing platoon size constraints. A callback function of the platoon size constraint is also feasible.
	\end{description}	
	
	This heuristic aims to encourage the formation of truck platoons and reduce their time flexibility. It affects the optimality of the scheduling problem but reduces the complexity of the problem. The more pairwise platoons are selected (the higher the ratio $\gamma$), the faster the heuristic, and the greater the optimality gap.

	\section{Computational results}\label{numericaldiscussion}
	
	To examine the computational performance of our models, we construct several instances based on two different networks; the data generation and implementation details are described in Section~\ref{IG}. In Section~\ref{MIPP} we first provide a comparison of the two monolithic MIP formulations CFP and TSF for TPP, followed by a comparison of heuristics in Section~\ref{HeurP}. %In Section~\ref{Management} we report our findings for a sensitivity analysis with which we try to discover under which settings the fuel savings are the largest.

	\subsection{Data generation and implementation details}\label{IG}

	\begin{figure}[h]
		\centering
		\begin{minipage}[t]{0.5\linewidth}
			\centering
			\raisebox{0.3\height}{\includegraphics[scale=0.3]{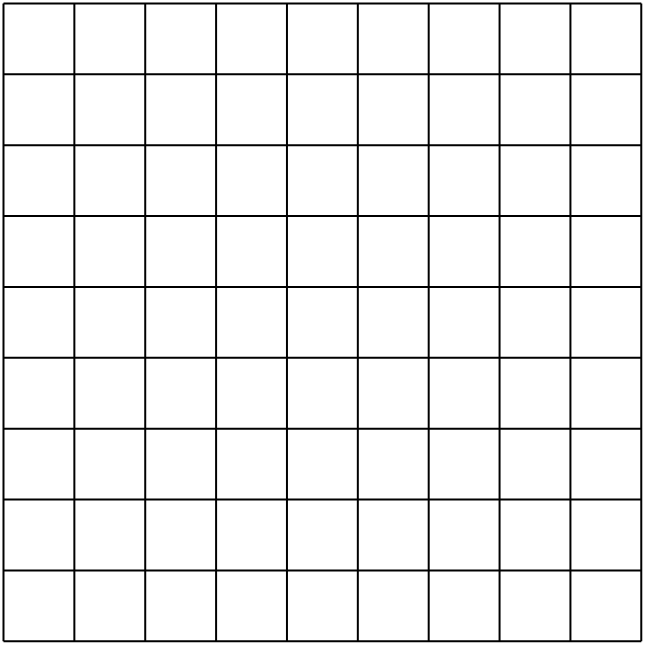}}
			\caption{Grid network}\label{grid}
		\end{minipage}%
		\begin{minipage}[t]{0.5\linewidth}
			\centering
			\includegraphics[scale=0.3]{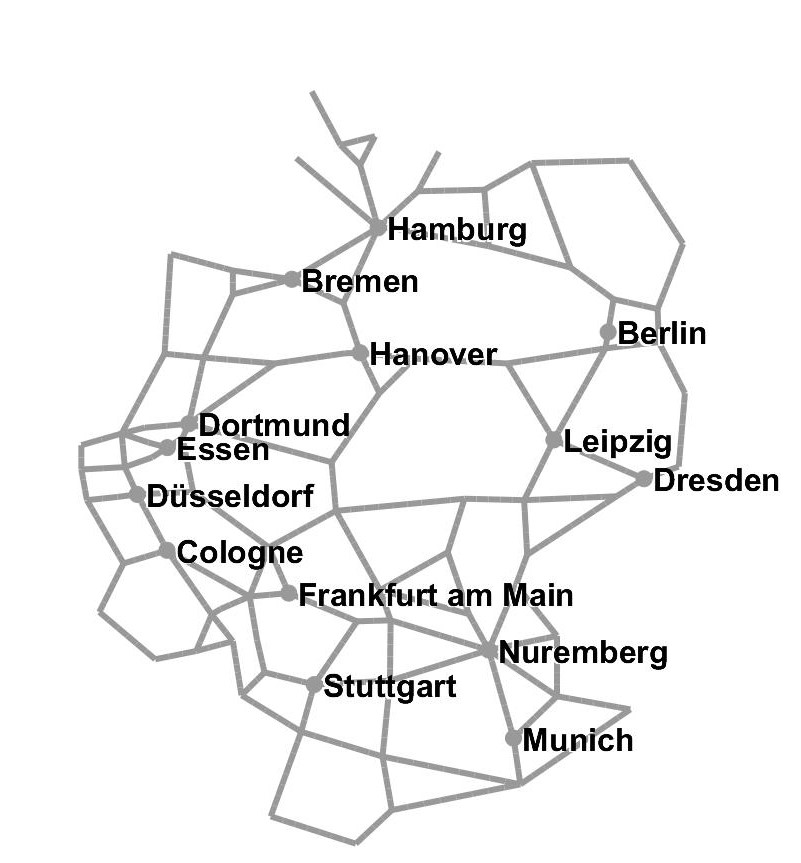}
			\caption{German highway network}\label{fig5}
		\end{minipage}
	\end{figure}

	In all our computations, the platooning fuel-saving rate $\eta$ is set as 10\%. We work with two different networks: a virtual $10 \times 10$ grid network (Fig.~\ref{grid}) and the German highway network from \citet{Luo2020} (Fig.~\ref{fig5}). The grid network consists of 100 nodes. For every pair of nodes that are adjacent in the grid, the distance is an integer value drawn randomly (and uniformly) from $\{3,4,5\}$.
	The German network consists of 647 nodes and 1390 arcs. To construct a time-space network, we discretize the time and distance parameters in the network. Here, we assume the speed on the German highway to be 120 km/h. Each period of time represents 10 minutes, and the total duration is within 24 hours. Since the time index may affect computation for different formulations of TPP and the scheduling problem, we also set different time period units (min) $TU\in \{10, 5, 1, 0.5\}$ for the German instances to investigate the effects.
	
	For the instances based on the grid network (Grid instances), we consider the following values for the total number of vehicles: $\{50, 100, 200, 400, 800, 1200, 1600\}$. %for the medium-sized instances we work with values in  $\{50, 100, 200, 400, 800\}$, while for large instances $|V|$ is in $\{1200, 1600\}$. 
	For the instances based on the German network (German instances), the total number of vehicles is also chosen from the set $\{30, 50, 100, 200, 300, 400, 600, 800\}$. % two different sets: medium-sized from  $\{100, 200, 300, $ $400, 500\}$ and large from $\{500, 600, 700, 800, 900, 1000\}$. 
	Each origin-destination (OD) pair is selected with uniform probability from among all node pairs for the grid network, while for the German network the OD pairs are generated using the distributed approach from \citet{Luo2020}, in which 75\% of the OD pairs are randomly generated from the nodes around the 14 major cities in Germany (within an hour drive), while 25\% of the OD pairs are uniformly generated from the entire node set.  In \citeauthor{Luo2020}'s generation routine, we use value two as the random seed. 
	The time windows are determined by the driving time and the earliest departure time. The allowed driving time of each vehicle is set to $1.2$ times the shortest driving time from its origin to destination. The earliest departure time is randomly generated within the first 12 hours of the time horizon.

	All our algorithms are implemented using the Python programming language. All computational experiments are run on a system with an Intel Core i7-7820HQ processor with 2.90 GHz CPU speed and 32 GB of RAM under a Windows 10 64-bit OS\@. All linear formulations are solved with the commercial solver Gurobi 9.0.3 with a single thread; all other Gurobi parameters are set to their default values. 
	For the  routing, scheduling, and truck platooning problems, several pre-processing techniques are applied to reduce the complexity of the problems, such as arc selection for vehicles (see Lemma~\ref{lem-1}) and arc combination \citep{Luo2020}.  For brevity, we will not explain these techniques in detail here.
	
	\begin{lemma}[\citealt{Larson2016}]
		If a following vehicle can save fraction $\eta$ of fuel consumption, then the path length of the vehicle should not exceed $1/(1-\eta)$ times  its shortest path length. Otherwise, the vehicle will waste more fuel on the extra distance.
		\label{lem-1}
	\end{lemma}
	
	\subsection{Performance of the MIP formulations for TPP}\label{MIPP}

	\begin{table}[h]
		\caption{Computational performance of the two MIP formulations for TPP}
		\scriptsize
		\setlength{\tabcolsep}{0.7mm}{
			\begin{threeparttable}[b]
				%\caption{Survey of bond lengths.\tnote{a}}
				%\label{Tab:bondlength}
				\begin{tabular}{@{}ccrrrrrrrrrrrrrrrrrr@{}}
					\toprule
					Size                                                                                           &                          & \multicolumn{6}{c}{$Q=4$}                                                                                                                                     & \multicolumn{6}{c}{$Q=5$}                                                                                                                                     & \multicolumn{6}{c}{$Q=6$}                                                                                                                                   \\ \midrule
					&                          & \multicolumn{3}{c}{CPF}                                                       & \multicolumn{3}{c}{TSF}                                                       & \multicolumn{3}{c}{CPF}                                                       & \multicolumn{3}{c}{TSF}                                                       & \multicolumn{3}{c}{CPF}                                                       & \multicolumn{3}{c}{TSF}                                                     \\ \midrule
					Net                                                                                            & V                        & \multicolumn{1}{c}{Gap} & \multicolumn{1}{c}{CPU} & \multicolumn{1}{c}{Sav}   & \multicolumn{1}{c}{Gap} & \multicolumn{1}{c}{CPU} & \multicolumn{1}{c}{Sav}   & \multicolumn{1}{c}{Gap} & \multicolumn{1}{c}{CPU} & \multicolumn{1}{c}{Sav}   & \multicolumn{1}{c}{Gap} & \multicolumn{1}{c}{CPU} & \multicolumn{1}{c}{Sav}   & \multicolumn{1}{c}{Gap} & \multicolumn{1}{c}{CPU} & \multicolumn{1}{c}{Sav}   & \multicolumn{1}{c}{Gap} & \multicolumn{1}{c}{CPU} & \multicolumn{1}{c}{Sav} \\ \midrule
					\multicolumn{1}{c|}{\multirow{5}{*}{Grid}}                                                     & \multicolumn{1}{c|}{50}  & 0.00                    & 0.1                     & \multicolumn{1}{r|}{0.95} & 0.00                    & 0.3                     & \multicolumn{1}{r|}{0.95} & 0.00                    & 0.1                     & \multicolumn{1}{r|}{0.95} & 0.00                    & 0.3                     & \multicolumn{1}{r|}{0.95} & 0.00                    & 0.1                     & \multicolumn{1}{r|}{0.95} & 0.00                    & 0.3                     & 0.95                    \\
					\multicolumn{1}{c|}{}                                                                          & \multicolumn{1}{c|}{100} & 0.00                    & 0.8                     & \multicolumn{1}{r|}{2.03} & 0.00                    & 0.6                     & \multicolumn{1}{r|}{2.03} & 0.00                    & 1.0                     & \multicolumn{1}{r|}{2.03} & 0.00                    & 0.7                     & \multicolumn{1}{r|}{2.03} & 0.00                    & 0.9                     & \multicolumn{1}{r|}{2.03} & 0.00                    & 0.6                     & 2.03                    \\
					\multicolumn{1}{c|}{}                                                                          & \multicolumn{1}{c|}{200} & 0.00                    & 170.8                   & \multicolumn{1}{r|}{3.07} & 0.00                    & 10.1                    & \multicolumn{1}{r|}{3.07} & 0.00                    & 79.2                    & \multicolumn{1}{r|}{3.09} & 0.00                    & 2.1                     & \multicolumn{1}{r|}{3.08} & 0.00                    & 60.1                    & \multicolumn{1}{r|}{3.09} & 0.00                    & 1.4                     & 3.10                    \\
					\multicolumn{1}{c|}{}                                                                          & \multicolumn{1}{c|}{400} & 0.11                    & 1800.0                  & \multicolumn{1}{r|}{4.05} & 0.04                    & 1800.0                  & \multicolumn{1}{r|}{4.07} & 0.05                    & 1800.0                  & \multicolumn{1}{r|}{4.13} & 0.02                    & 17.8                    & \multicolumn{1}{r|}{4.13} & 0.05                    & 1800.0                  & \multicolumn{1}{r|}{4.14} & 0.01                    & 5.5                     & 4.15                    \\
					\multicolumn{1}{c|}{}                                                                          & \multicolumn{1}{c|}{800} & 2.06                    & 1800.0                  & \multicolumn{1}{r|}{3.95} & \multicolumn{3}{c|}{no}                                                       & 1.79                    & 1800.0                  & \multicolumn{1}{r|}{4.16} & \multicolumn{3}{c|}{no}                                                       & 1.87                    & 1800.0                  & \multicolumn{1}{r|}{4.21} & \multicolumn{3}{c}{no}                                                      \\ \midrule
					\multicolumn{1}{c|}{\multirow{5}{*}{\begin{tabular}[c]{@{}c@{}}Ger-\\      many\end{tabular}}} & \multicolumn{1}{c|}{30}  & 0.00                    & 0.1                     & \multicolumn{1}{r|}{1.49} & 0.00                    & 0.3                     & \multicolumn{1}{r|}{1.49} & 0.00                    & 0.2                     & \multicolumn{1}{r|}{1.49} & 0.00                    & 0.3                     & \multicolumn{1}{r|}{1.49} & 0.00                    & 0.1                     & \multicolumn{1}{r|}{1.49} & 0.00                    & 0.3                     & 1.49                    \\
					\multicolumn{1}{c|}{}                                                                          & \multicolumn{1}{c|}{50}  & 0.00                    & 0.4                     & \multicolumn{1}{r|}{2.06} & 0.00                    & 0.7                     & \multicolumn{1}{r|}{2.06} & 0.00                    & 0.4                     & \multicolumn{1}{r|}{2.06} & 0.00                    & 0.8                     & \multicolumn{1}{r|}{2.06} & 0.00                    & 0.4                     & \multicolumn{1}{r|}{2.06} & 0.00                    & 0.7                     & 2.06                    \\
					\multicolumn{1}{c|}{}                                                                          & \multicolumn{1}{c|}{100} & 0.00                    & 4.1                     & \multicolumn{1}{r|}{2.63} & 0.00                    & 1.9                     & \multicolumn{1}{r|}{2.63} & 0.00                    & 3.5                     & \multicolumn{1}{r|}{2.63} & 0.00                    & 2.0                     & \multicolumn{1}{r|}{2.63} & 0.00                    & 4.3                     & \multicolumn{1}{r|}{2.63} & 0.00                    & 1.9                     & 2.63                    \\
					\multicolumn{1}{c|}{}                                                                          & \multicolumn{1}{c|}{200} & 0.03                    & 1800.0                  & \multicolumn{1}{r|}{4.16} & 0.04                    & 1800.0                  & \multicolumn{1}{r|}{4.16} & 0.01                    & 1800.0                  & \multicolumn{1}{r|}{4.21} & 0.03                    & 1800.0                  & \multicolumn{1}{r|}{4.20} & 0.01                    & 1800.0                  & \multicolumn{1}{r|}{4.27} & 0.00                    & 25.1                    & 4.27                    \\
					\multicolumn{1}{c|}{}                                                                          & \multicolumn{1}{c|}{400} & 1.77                    & 1800.0                  & \multicolumn{1}{r|}{3.95} & \multicolumn{3}{c|}{no}                                                       & 1.33                    & 1800.0                  & \multicolumn{1}{r|}{4.28} & \multicolumn{3}{c|}{no}                                                       & 1.25                    & 1800.0                  & \multicolumn{1}{r|}{4.51} & \multicolumn{3}{c}{no}                                                      \\ \bottomrule
				\end{tabular}
				\begin{tablenotes}   
					\scriptsize    
					\item[1] Q: platoon size limit; Net: Network type; V: vehicle number; Gap means relative gap (\%) within 30 min;  
					CPU: CPU time (s); Sav: saving ratio (\%); ``no" means that no feasible solution is found within 30 minutes.
				\end{tablenotes}
		\end{threeparttable}}%
		\label{table:entirePlatoon}
	\end{table}

\begin{table}[h]
\caption{Computational performance for TPP with German instances ($Q=5$) under different time indices}
\scriptsize
\centering
%\resizebox{\textwidth}{!}{ %
	\setlength{\tabcolsep}{1.0mm}{
		\begin{threeparttable}[b]
			%\caption{Survey of bond lengths.\tnote{a}}
			%\label{Tab:bondlength}
			\begin{tabular}{crrrrrrrrrrrrrrrr}
				\hline
				V   & \multicolumn{4}{c}{10}                                                                                  & \multicolumn{4}{c}{30}                                                                                   & \multicolumn{4}{c}{50}                                                                                   & \multicolumn{4}{c}{100}                                                                               \\ \hline
				& \multicolumn{2}{c}{CPF}                           & \multicolumn{2}{c}{TSF}                             & \multicolumn{2}{c}{CPF}                           & \multicolumn{2}{c}{TSF}                              & \multicolumn{2}{c}{CPF}                           & \multicolumn{2}{c}{TSF}                              & \multicolumn{2}{c}{CPF}                           & \multicolumn{2}{c}{TSF}                           \\ \cline{2-17} 
				TU  & \multicolumn{1}{c}{Gap} & \multicolumn{1}{c}{CPU} & \multicolumn{1}{c}{Gap} & \multicolumn{1}{c}{CPU}   & \multicolumn{1}{c}{Gap} & \multicolumn{1}{c}{CPU} & \multicolumn{1}{c}{Gap} & \multicolumn{1}{c}{CPU}    & \multicolumn{1}{c}{Gap} & \multicolumn{1}{c}{CPU} & \multicolumn{1}{c}{Gap} & \multicolumn{1}{c}{CPU}    & \multicolumn{1}{c}{Gap} & \multicolumn{1}{c}{CPU} & \multicolumn{1}{c}{Gap} & \multicolumn{1}{c}{CPU} \\ \hline
				10  & 0.00                    & 0.1                     & 0.00                    & \multicolumn{1}{r|}{0.3}  & 0.00                    & 0.2                     & 0.00                    & \multicolumn{1}{r|}{0.3}   & 0.00                    & 0.4                     & 0.00                    & \multicolumn{1}{r|}{0.8}   & 0.00                    & 3.5                     & 0.00                    & 2.0                     \\
				5   & 0.00                    & \textless{}0.1          & 0.00                    & \multicolumn{1}{r|}{0.4}  & 0.00                    & 3.2                     & 0.00                    & \multicolumn{1}{r|}{2.2}   & 0.00                    & 165.0                   & 0.00                    & \multicolumn{1}{r|}{7.6}   & 0.66                    & 1800.0                  & 0.00                    & 54.2                    \\
				1   & 0.00                    & 0.1                     & 0.00                    & \multicolumn{1}{r|}{15.4} & 0.00                    & 73.0                    & 0.00                    & \multicolumn{1}{r|}{37.2}  & 1.02                    & 1800.0                  & 0.00                    & \multicolumn{1}{r|}{165.3} & 1.05                    & 1800.0                  & 0.00                    & 518.0                   \\
				0.5 & 0.00                    & 0.2                     & 0.00                    & \multicolumn{1}{r|}{23.1} & 0.00                    & 89.1                    & 0.00                    & \multicolumn{1}{r|}{153.9} & 1.05                    & 1800.0                  & 0.00                    & \multicolumn{1}{r|}{577.3} & 1.07                    & 1800.0                  & 0.00                    & 1614.7                  \\ \hline
			\end{tabular}
			\begin{tablenotes}   
				\scriptsize         
				\item[1] TU indicates the minimum time unit (min); Gap means the relative gap (\%) within 30 min; CPU: CPU time (s). %``no" means that no feasible solution is found within 30 minutes.
			\end{tablenotes}
	\end{threeparttable}}%
\label{table:PlatoonT}
\end{table}

	Table~\ref{table:entirePlatoon} and \ref{table:PlatoonT} contain the results of a comparison of the two MIP formulations, CPF and TSF, for TPP\@. Table~\ref{table:PlatoonT} only examines lower value for vehicle number $V$, as it is sufficient for comparisons, and for other instances with a larger number of vehicles, TSF was unable to obtain any feasible solution within 30 minutes.
	We report the following indicators: the saving ratio, the relative gap when the formulation is truncated after 30 minutes of runtime, and the CPU time for the formulation to find a guaranteed optimal solution. 
	The saving ratio indicates the cost savings that the platooning mode can achieve compared to the shortest-path cost without platooning. Here, the cost savings is equal to the shortest path cost minus the best-found objective value.
	The relative gap is computed as the percentage difference between the best-found solution value and the best-known bound within the time limit. %\textcolor{blue}{A summary of indicators is also shown in Table~\ref{table:indicatornotation}}.
	
	The comparison of TSF and CPF is influenced by various factors. In terms of quantity, based on Table~\ref{table:entirePlatoon}, when the number of vehicles is limited (e.g., German instances with 30 and 50 vehicles), CPF requires less CPU time than TSF to achieve optimality. However, when the number of vehicles is large (e.g., German instances with 200 vehicles, and Grid instances with 200 and 400 vehicles), TSF requires less CPU time or achieves a smaller gap within the 30-minute time limit than CPF. On the other hand, for very large vehicle numbers (e.g., the Grid instance with 800 vehicles and the German instance with 400 vehicles), CPF was able to find a good feasible solution within 30 minutes, while TSF was unable to find any feasible solution within that timeframe due to a long time spent on the root simplex.
	
	In terms of the platoon size limit $Q$, TSF is more likely to perform better in terms of time or quality when the limit is higher according to Table~\ref{table:entirePlatoon}. The row of the German instance with 200 vehicles can prove the insight. In detail, when $Q=4$, TSF performs worse than CPF due to a higher relative gap, while when $Q=6$, TSF performs much better due to a lower CPU time required to reach optimality. 
	
	In terms of time unit $TU$, as its accuracy increases ($TU$ decreases), both CPF and TSF suffer in terms of time and solution quality, but TSF is more affected according to Table~\ref{table:PlatoonT}. Additionally, for CPF, if the time unit is already relatively low, the impact of time precision may be limited. For example, for the German instances, when the time unit changes from 10 minutes to 5 minutes, the CPU time required by CPF for reaching optimality increases significantly; and when the time unit changes from 1 minute to 0.5 minutes, the CPU time required does not change significantly and the gap remains almost the same.
	
	Mathematically, the number of decision variables for CPF is $O(|V|^2|M|)$, where $|V|$ is the number of vehicles and $|M|$ is the number of arcs. The number of decision variables for TSF is $O(\beta |V||M||T|)$, where $|T|$ is the time unit number and $\beta$ is the maximum number of truck platoons on an arc at a time, $1\le \beta \le |V|/Q$. The variable numbers explain the advantage of CPF on time and TIF on vehicle numbers. For platoon size limit, when $Q$ is high, $\beta$ tends to be 1, hence TSF can have a lower number of variables and can  quickly lead to an optimal solution.
	
	In addition to these results regarding the comparison of the formulations, we observe that more vehicles and higher platoon size limits are beneficial for platooning savings. 
	
	\subsection{Performance of the heuristics for TPP}\label{HeurP}

	To show the effectiveness of our variant of the decomposition-based platooning heuristic, we compare the performance with the implementation of  \citet{Luo2020}, who also developed the overall algorithmic framework. Other heuristics for TPP have been published, but we do not directly compare with those. 
	
	\citet{ABDOLMALEKI202191} propose a dynamic-programming-based heuristic for TPP, but they aim to obtain a feasible solution quickly instead of a near-optimal one. \citet{Nourmohammadzadeh2019} study a similar problem setting except for the platoon size limit. After communication with the authors it turns out that the original instances are not available. For their settings without considering the size limit, we do note that  our heuristic can achieve an optimality gap from 0.01\% to 0.05\% for similar-sized instances (see Table~\ref{table:iterative} in the appendix), which is significantly lower than their approach. Here, the optimality gap is the percentage by which the objective value of the best-found solution exceeds the optimal objective value.
	
	The comparisons with \citet{Luo2020} are based on three aspects, namely the routing problem (see Section~\ref{subsub:compare-routing}), the scheduling problem (Section~\ref{subsub:compare-scheduling}), and TPP (Section~\ref{subsub:compare-overall}). For TPP, the comparisons involve both the iterative process (the cost-modification procedure) and the full heuristic. 
	
	\begin{table}[h]
		\footnotesize   
		\centering
		\caption{Comparisons in Section~\ref{HeurP} }
		\begin{threeparttable}
			\begin{tabular}{@{}lll@{}}
				\toprule
				& \citet{Luo2020}   & \emph{This paper}               \\\midrule
				Routing problem           & LRF              & FCNF               \\\midrule
				Scheduling problem        & ABF              & TIF                \\\midrule
				\multirow{2}{*}{\begin{tabular}[l]{@{}c@{}}Iterative process\\  and heuristics   \end{tabular}} & LLIter (FCNF + TIF + LLCMP) \tnote{1}  & \multirow{2}{*}{IHeur (FCNF + TIF + ICMP)\tnote{3}}   \\%\cmidrule(r){1-2}
				        & LLHeur (LRF + ABF + LLCMP) \tnote{2} &    \\\bottomrule
			\end{tabular}
					\begin{tablenotes}   
			\footnotesize               
			\item[1] LLIter: \citeauthor{Luo2020}'s iteration. 
			\item[2] LLHeur: \citeauthor{Luo2020}'s heuristic.
			\item[3] IHeur: our improved heuristic.
		\end{tablenotes}
		\end{threeparttable} 
		\label{table:comparisonall}
	\end{table}

	An overview of the settings that will be compared  is presented in Table~\ref{table:comparisonall}; we refer to Table~\ref{table:fnotation} and Table~\ref{table:indicatornotation} in the appendix for a summary of the abbreviations and indicators.
	
	\subsubsection{Performance of formulations for the routing problem} \label{subsub:compare-routing}
	
	\begin{table}[h]
		\footnotesize
		\centering
		\caption{CPU time (s) for solving the routing problem}
		\begin{threeparttable}
			\begin{tabular}{@{}cccccc@{}}
				\toprule
				\multicolumn{3}{c}{Grid network}                                                          & \multicolumn{3}{c}{German network}                                                        \\ \midrule
				\multicolumn{1}{l}{Vehicles} & \multicolumn{1}{l}{LRF} & \multicolumn{1}{l}{FCNF} & \multicolumn{1}{l}{Vehicles} & \multicolumn{1}{l}{LRF} & \multicolumn{1}{l}{FCNF} \\\midrule
				200                          & 0.14                    & 0.13                     & 200                          & 0.39                    & 0.19                     \\
				400                          & 0.28                    & 0.23                     & 400                          & 0.85                    & 0.59                     \\
				800                          & 0.41                    & 0.45                     & 800                          & 1.71                    & 0.93                     \\
				1000                         & 0.46                    & 0.46                     & 1000                         & 3.48                    & 2.42                     \\
				2000                         & 2.46                    & 1.24                     & 2000                         & 7.31                    & 5.84                     \\
				4000                         & 2.51                    & 1.92                     & 4000                         & 11.30                   & 7.23                     \\
				8000                         & 4.18                    & 2.96                     & 8000                         & 24.94                   & 13.58                    \\ \bottomrule
			\end{tabular}
		\end{threeparttable} 
		\label{table:routingProblem}
	\end{table}
	
	Table~\ref{table:routingProblem} shows the computational performance of the two alternative formulations LRF and FCNF for the routing problem, as described in Section~\ref{routeP}. Both models run quite fast, and we therefore also test  very large instances to distinguish the two formulations. Overall, we conclude that FCNF is faster than LRF.
	
	\subsubsection{Performance of formulations for the scheduling problem} \label{subsub:compare-scheduling}

\begin{table}[h]
\centering
\caption{Computational performance for the scheduling problem with different platoon size limits ($TU=10$)}\label{schedulingResult}
\scriptsize
%\resizebox{\textwidth}{!}{ %
	\setlength{\tabcolsep}{0.5mm}{
		\begin{threeparttable}[b]
			%\caption{Survey of bond lengths.\tnote{a}}
			%\label{Tab:bondlength}
			\begin{tabular}{ccrrrrrrrrrrrrrrrrrr}
				\hline
				Size                                                                      & \multicolumn{1}{l}{} & \multicolumn{6}{c}{$Q=4$}                                                                                                                                                                                                                                                                                                                                                                   & \multicolumn{6}{c}{$Q=5$}                                                                                                                                                                                                                                                                                                                                                                   & \multicolumn{6}{c}{$Q=6$}                                                                                                                                                                                                                                                                                                                                                               \\ \hline
				\multicolumn{1}{l}{}                                                      & \multicolumn{1}{l}{} & \multicolumn{3}{c}{ABF}                                                                                                                                                                         & \multicolumn{3}{c}{TIF}                                                                                                                                                                 & \multicolumn{3}{c}{ABF}                                                                                                                                                                         & \multicolumn{3}{c}{TIF}                                                                                                                                                                 & \multicolumn{3}{c}{ABF}                                                                                                                                                                         & \multicolumn{3}{c}{TIF}                                                                                                                                                             \\ \hline
				Net                                                                       & V                    & \multicolumn{1}{c}{\begin{tabular}[c]{@{}c@{}}Gap\\      (10)\end{tabular}} & \multicolumn{1}{c}{\begin{tabular}[c]{@{}c@{}}Gap\\      (30)\end{tabular}} & \multicolumn{1}{c}{CPU}             & \multicolumn{1}{c}{\begin{tabular}[c]{@{}c@{}}Gap\\      (10)\end{tabular}} & \multicolumn{1}{c}{\begin{tabular}[c]{@{}c@{}}Gap\\      (30)\end{tabular}} & \multicolumn{1}{c}{CPU}     & \multicolumn{1}{c}{\begin{tabular}[c]{@{}c@{}}Gap\\      (10)\end{tabular}} & \multicolumn{1}{c}{\begin{tabular}[c]{@{}c@{}}Gap\\      (30)\end{tabular}} & \multicolumn{1}{c}{CPU}             & \multicolumn{1}{c}{\begin{tabular}[c]{@{}c@{}}Gap\\      (10)\end{tabular}} & \multicolumn{1}{c}{\begin{tabular}[c]{@{}c@{}}Gap\\      (30)\end{tabular}} & \multicolumn{1}{c}{CPU}     & \multicolumn{1}{c}{\begin{tabular}[c]{@{}c@{}}Gap\\      (10)\end{tabular}} & \multicolumn{1}{c}{\begin{tabular}[c]{@{}c@{}}Gap\\      (30)\end{tabular}} & \multicolumn{1}{c}{CPU}             & \multicolumn{1}{c}{\begin{tabular}[c]{@{}c@{}}Gap\\      (10)\end{tabular}} & \multicolumn{1}{c}{\begin{tabular}[c]{@{}c@{}}Gap\\      (30)\end{tabular}} & \multicolumn{1}{c}{CPU} \\ \hline
				\multirow{5}{*}{Grid}                                                     & 100                  & 0.00                                                                        & 0.00                                                                        & \multicolumn{1}{r|}{\textless{}0.1} & 0.00                                                                        & 0.00                                                                        & \multicolumn{1}{r|}{0.1}    & 0.00                                                                        & 0.00                                                                        & \multicolumn{1}{r|}{\textless{}0.1} & 0.00                                                                        & 0.00                                                                        & \multicolumn{1}{r|}{0.1}    & 0.00                                                                        & 0.00                                                                        & \multicolumn{1}{r|}{\textless{}0.1} & 0.00                                                                        & 0.00                                                                        & 0.1                     \\
				& 200                  & 0.00                                                                        & 0.00                                                                        & \multicolumn{1}{r|}{0.9}            & 0.00                                                                        & 0.00                                                                        & \multicolumn{1}{r|}{0.5}    & 0.00                                                                        & 0.00                                                                        & \multicolumn{1}{r|}{0.4}            & 0.00                                                                        & 0.00                                                                        & \multicolumn{1}{r|}{0.2}    & 0.00                                                                        & 0.00                                                                        & \multicolumn{1}{r|}{0.5}            & 0.00                                                                        & 0.00                                                                        & 0.2                     \\
				& 400                  & 0.17                                                                        & 0.06                                                                        & \multicolumn{1}{r|}{1800.0}         & 0.23                                                                        & 0.00                                                                        & \multicolumn{1}{r|}{1125.6} & 0.00                                                                        & 0.00                                                                        & \multicolumn{1}{r|}{67.2}           & 0.00                                                                        & 0.00                                                                        & \multicolumn{1}{r|}{2.3}    & 0.00                                                                        & 0.00                                                                        & \multicolumn{1}{r|}{89.2}           & 0.00                                                                        & 0.00                                                                        & 1.0                     \\
				& 600                  & 2.06                                                                        & 1.68                                                                        & \multicolumn{1}{r|}{1800.0}         & 1.60                                                                        & 1.25                                                                        & \multicolumn{1}{r|}{1800.0} & 1.71                                                                        & 1.44                                                                        & \multicolumn{1}{r|}{1800.0}         & 0.87                                                                        & 0.70                                                                        & \multicolumn{1}{r|}{1800.0} & 1.05                                                                        & 0.82                                                                        & \multicolumn{1}{r|}{1800.0}         & 0.00                                                                        & 0.00                                                                        & 507.9                   \\
				& 800                  & 16.55                                                                       & 5.91                                                                        & \multicolumn{1}{r|}{1800.0}         & 4.52                                                                        & 3.06                                                                        & \multicolumn{1}{r|}{1800.0} & 17.15                                                                       & 3.85                                                                        & \multicolumn{1}{r|}{1800.0}         & 1.70                                                                        & 1.39                                                                        & \multicolumn{1}{r|}{1800.0} & 3.65                                                                        & 2.80                                                                        & \multicolumn{1}{r|}{1800.0}         & 0.68                                                                        & 0.56                                                                        & 1800.0                  \\ \hline
				\multirow{5}{*}{\begin{tabular}[c]{@{}c@{}}Ger-\\      many\end{tabular}} & 50                   & 0.00                                                                        & 0.00                                                                        & \multicolumn{1}{r|}{\textless{}0.1} & 0.00                                                                        & 0.00                                                                        & \multicolumn{1}{r|}{0.1}    & 0.00                                                                        & 0.00                                                                        & \multicolumn{1}{r|}{\textless{}0.1} & 0.00                                                                        & 0.00                                                                        & \multicolumn{1}{r|}{0.1}    & 0.00                                                                        & 0.00                                                                        & \multicolumn{1}{r|}{\textless{}0.1} & 0.00                                                                        & 0.00                                                                        & 0.1                     \\
				& 100                  & 0.00                                                                        & 0.00                                                                        & \multicolumn{1}{r|}{0.1}            & 0.00                                                                        & 0.00                                                                        & \multicolumn{1}{r|}{0.8}    & 0.00                                                                        & 0.00                                                                        & \multicolumn{1}{r|}{0.1}            & 0.00                                                                        & 0.00                                                                        & \multicolumn{1}{r|}{0.2}    & 0.00                                                                        & 0.00                                                                        & \multicolumn{1}{r|}{0.1}            & 0.00                                                                        & 0.00                                                                        & 0.3                     \\
				& 200                  & 0.00                                                                        & 0.00                                                                        & \multicolumn{1}{r|}{124.3}          & 0.60                                                                        & 0.42                                                                        & \multicolumn{1}{r|}{1800.0} & 0.00                                                                        & 0.00                                                                        & \multicolumn{1}{r|}{128.2}          & 0.60                                                                        & 0.17                                                                        & \multicolumn{1}{r|}{1800.0} & 0.00                                                                        & 0.00                                                                        & \multicolumn{1}{r|}{9.7}            & 0.00                                                                        & 0.00                                                                        & 7.5                     \\
				& 300                  & 1.18                                                                        & 0.98                                                                        & \multicolumn{1}{r|}{1800.0}         & 1.07                                                                        & 0.87                                                                        & \multicolumn{1}{r|}{1800.0} & 1.03                                                                        & 0.88                                                                        & \multicolumn{1}{r|}{1800.0}         & 0.49                                                                        & 0.44                                                                        & \multicolumn{1}{r|}{1800.0} & 0.44                                                                        & 0.35                                                                        & \multicolumn{1}{r|}{1800.0}         & 0.11                                                                        & 0.07                                                                        & 1800.0                  \\
				& 400                  & 20.54                                                                       & 12.48                                                                       & \multicolumn{1}{r|}{1800.0}         & no                                                                          & 2.74                                                                        & \multicolumn{1}{r|}{1800.0} & 15.85                                                                       & 8.97                                                                        & \multicolumn{1}{r|}{1800.0}         & 1.67                                                                        & 1.10                                                                        & \multicolumn{1}{r|}{1800.0} & 17.33                                                                       & 9.99                                                                        & \multicolumn{1}{r|}{1800.0}         & 0.47                                                                        & 0.37                                                                        & 1800.0                  \\ \hline
			\end{tabular}
			\begin{tablenotes}   
				\scriptsize         
				\item[1] Size $Q$: platoon size limit; Net: Network type; V: vehicle number; Gap (10) means relative gap (\%) within 10 minutes; Gap (30) means relative gap  (\%) in 30 minutes; CPU: CPU time (s). ``no" means that no feasible solution is found within a certain time limit.
			\end{tablenotes}
	\end{threeparttable}}%
\label{table:schedulingProblem}
\end{table}

\begin{table}[h]
\centering
\caption{Computational performance for the scheduling problem ($Q=5$) under different time units within 30 minutes}\label{schedulingT}
\scriptsize
\setlength{\tabcolsep}{1mm}{
\begin{threeparttable}[b]
\begin{tabular}{crrrrrrrrrrcrrrcr}
	\hline
	V   & \multicolumn{4}{c}{50}                                                                                                                                                                                                                                                                                              & \multicolumn{4}{c}{100}                                                                                                                                                                                                                                                                                             & \multicolumn{4}{c}{200}                                                                                                                                                                                                                                                                         & \multicolumn{4}{c}{300}                                                                                                                                                                                                                                                                         \\ \hline
	& \multicolumn{2}{c}{ABF}                                                                                                                                  & \multicolumn{2}{c}{TIF}                                                                                                                                  & \multicolumn{2}{c}{ABF}                                                                                                                                  & \multicolumn{2}{c}{TIF}                                                                                                                                  & \multicolumn{2}{c}{ABF}                                                                                                                                  & \multicolumn{2}{c}{TIF}                                                                                                              & \multicolumn{2}{c}{ABF}                                                                                                                                  & \multicolumn{2}{c}{TIF}                                                                                                              \\ \cline{2-17} 
	TU  & \multicolumn{1}{c}{\begin{tabular}[c]{@{}c@{}}Gap\\      (\%)\end{tabular}} & \multicolumn{1}{c}{\begin{tabular}[c]{@{}c@{}}CPU\\      (s)\end{tabular}} & \multicolumn{1}{c}{\begin{tabular}[c]{@{}c@{}}Gap\\      (\%)\end{tabular}} & \multicolumn{1}{c}{\begin{tabular}[c]{@{}c@{}}CPU\\      (s)\end{tabular}} & \multicolumn{1}{c}{\begin{tabular}[c]{@{}c@{}}Gap\\      (\%)\end{tabular}} & \multicolumn{1}{c}{\begin{tabular}[c]{@{}c@{}}CPU\\      (s)\end{tabular}} & \multicolumn{1}{c}{\begin{tabular}[c]{@{}c@{}}Gap\\      (\%)\end{tabular}} & \multicolumn{1}{c}{\begin{tabular}[c]{@{}c@{}}CPU\\      (s)\end{tabular}} & \multicolumn{1}{c}{\begin{tabular}[c]{@{}c@{}}Gap\\      (\%)\end{tabular}} & \multicolumn{1}{c}{\begin{tabular}[c]{@{}c@{}}CPU\\      (s)\end{tabular}} & \begin{tabular}[c]{@{}c@{}}Gap\\      (\%)\end{tabular} & \multicolumn{1}{c}{\begin{tabular}[c]{@{}c@{}}CPU\\      (s)\end{tabular}} & \multicolumn{1}{c}{\begin{tabular}[c]{@{}c@{}}Gap\\      (\%)\end{tabular}} & \multicolumn{1}{c}{\begin{tabular}[c]{@{}c@{}}CPU\\      (s)\end{tabular}} & \begin{tabular}[c]{@{}c@{}}Gap\\      (\%)\end{tabular} & \multicolumn{1}{c}{\begin{tabular}[c]{@{}c@{}}CPU\\      (s)\end{tabular}} \\ \hline
	10  & 0.00                                                                        & \textless{}0.1                                                             & 0.00                                                                        & \multicolumn{1}{r|}{0.1}                                                   & 0.00                                                                        & 0.1                                                                        & 0.00                                                                        & \multicolumn{1}{r|}{0.2}                                                   & 0.00                                                                        & 128.2                                                                      & \multicolumn{1}{r}{0.17}                                & \multicolumn{1}{r|}{1800.0}                                                & 0.88                                                                        & 1800.0                                                                     & \multicolumn{1}{r}{0.44}                                & 1800.0                                                                     \\
	5   & 0.00                                                                        & \textless{}0.1                                                             & 0.00                                                                        & \multicolumn{1}{r|}{0.3}                                                   & 0.00                                                                        & 0.2                                                                        & 0.00                                                                        & \multicolumn{1}{r|}{1.2}                                                   & 0.00                                                                        & 48.4                                                                       & \multicolumn{1}{r}{0.00}                                & \multicolumn{1}{r|}{263.8}                                                 & 1.46                                                                        & 1800.0                                                                     & \multicolumn{1}{r}{0.64}                                & 1800.0                                                                     \\
	1   & 0.00                                                                        & \textless{}0.1                                                             & 0.00                                                                        & \multicolumn{1}{r|}{11.4}                                                  & 0.00                                                                        & 0.3                                                                        & 0.00                                                                        & \multicolumn{1}{r|}{23.2}                                                  & 0.00                                                                        & 161.9                                                                      & \multicolumn{1}{r}{0.78}                                & \multicolumn{1}{r|}{1800.0}                                                & 0.40                                                                        & 1800.0                                                                     & \multicolumn{2}{c}{no}                                                                                                               \\
	0.5 & 0.00                                                                        & \textless{}0.1                                                             & 0.00                                                                        & \multicolumn{1}{r|}{55.5}                                                  & 0.00                                                                        & 0.1                                                                        & 0.00                                                                        & \multicolumn{1}{r|}{105.7}                                                 & 0.00                                                                        & 112.6                                                                      & \multicolumn{2}{c|}{no}                                                                                                              & 0.93                                                                        & 1800.0                                                                     & \multicolumn{2}{c}{no}                                                                                                               \\ \hline
\end{tabular}
\begin{tablenotes}   
\scriptsize         
\item[1] V: vehicle number; TU indicates the minimum time unit (min); Gap means relative gap (\%) within 30 minutes. ``no" means that no feasible solution is found within 30 minutes.
\end{tablenotes}
\end{threeparttable}}%
\label{table:schedulingT}
\end{table}

	Our comparison of different formulations for the scheduling problem is based on the relative gap and CPU time (s). 
	
	Table~\ref{table:schedulingProblem} reports the results for instances with a different platoon size limit and the time unit of 10 minutes, and Table~\ref{table:schedulingT} presents the results for German instances with a specified platoon size limit ($Q=5$) but with varying time units. The time limit for running is 30 minutes.
	
	Table~\ref{table:schedulingProblem} and \ref{table:schedulingT} reveal that vehicle number, platoon size limit, and time unit can affect computational performance. Regarding the vehicle number, based on Table~\ref{table:schedulingProblem}, we find that TIF performs worse than ABF when the number of vehicles is small due to longer CPU time to get an optimal solution (e.g., the German instance with 100 vehicles). However, TIF can show computational advantages over ABF as the vehicle number increases because TIF can lead to a lower relative gap within the time framework (e.g. German instance with 300 vehicles). Moreover, when the vehicle number is very large, TIF cannot provide a feasible solution in a limited running time (e.g., when $Q = 4$, TIF cannot find a feasible solution for the German instances with 400 vehicles). As for the platoon size limit $Q$, from Table~\ref{table:schedulingProblem}, we see TIF is more likely to outperform ABF when the $Q$ is high. For example, in the row of the German network with 200 vehicles, TIF requires more CPU time than ABF when $Q=4$, but less time than ABF when $Q=6$. Table~\ref{table:schedulingT} provides insights of the impact of the time unit's choice. A direct insight is that TIF performs worse than ABF as the time unit becomes small. Nevertheless, as the vehicle number increases, TIF can still show advantages (e.g. when time unit $TU$ is 5 and vehicle number $V$ is 300).
	
	The comparison between ABF and TIF for the scheduling problem is similar to the comparison between CPF and TSF for TPP. The number of decision variables for ABF is $O(|V|^2|M|)$, while that for TIF is $O(\beta |V||M||T|)$, where $|V| $ is the number of vehicles, $|M|$ is the number of arcs, $|T|$ is the number of time units, and $\beta$ is the maximum number of truck platoons on an arc at a given time, with $1\le \beta \le |V|/Q$. Compared with CPF, TSF can take advantage in computation when the vehicle number is high, while it suffers when the number of time units is high (high time accuracy). For platoon size limit, when $Q$ is high, $\beta$ tends to be 1, so TSF has a lower number of variables, and using TSF for solver can achieve optimality easily. 
	
	ABF and TIF have their own advantages, but as long as the time accuracy keeps at a reasonable level (10 minutes), TIF can perform better than ABF for large instances. Otherwise, under the TPP heuristic framework, ABF might be used to replace TIF.

	\subsubsection{Performance of solution procedures for TPP} \label{subsub:compare-overall}

	\begin{table}[h]
		\footnotesize 
		\centering
		\caption{Computational performance of different solution methods for TPP ($Q=5$, $TU=10$)}
		\begin{threeparttable}
			\begin{tabular}{@{}ccrrcrrrrrr@{}}
				\toprule
				\multicolumn{1}{l}{}                                                      & \multicolumn{1}{l}{}  & \multicolumn{2}{c}{\begin{tabular}[c]{@{}c@{}}CPF\\      (10h)\end{tabular}}                                                                                          & \multicolumn{1}{l}{\begin{tabular}[c]{@{}l@{}}LLHeur\\      (30 min)\end{tabular}} & \multicolumn{3}{c}{\begin{tabular}[c]{@{}c@{}}LLIter\\      (30 min)\end{tabular}}                                                                                                                                                           & \multicolumn{3}{c}{\begin{tabular}[c]{@{}c@{}}IHeur\\       (30 min)\end{tabular}}                                                                                                                                       \\ \midrule
				\multicolumn{1}{l}{Net}                                                   & \multicolumn{1}{l}{V} & \multicolumn{1}{c}{\begin{tabular}[c]{@{}c@{}}UB sav\\       (\%)\end{tabular}} & \multicolumn{1}{c}{\begin{tabular}[c]{@{}c@{}}Sav\\      (\%)\end{tabular}} & \begin{tabular}[c]{@{}c@{}}Sav\\      (\%)\end{tabular}                             & \multicolumn{1}{c}{\begin{tabular}[c]{@{}c@{}}Best\\      Iter\end{tabular}} & \multicolumn{1}{c}{\begin{tabular}[c]{@{}c@{}}Total\\       Iter\end{tabular}} & \multicolumn{1}{c}{\begin{tabular}[c]{@{}c@{}}Sav\\      (\%)\end{tabular}} & \multicolumn{1}{c}{\begin{tabular}[c]{@{}c@{}}Best\\      Iter\end{tabular}} & \multicolumn{1}{c}{\begin{tabular}[c]{@{}c@{}}Total\\       Iter\end{tabular}} & \multicolumn{1}{c}{\begin{tabular}[c]{@{}c@{}}Sav\\      (\%)\end{tabular}} \\ \midrule
				\multirow{6}{*}{Grid}                                                     & 100                   & 2.03                                                                                    & 2.03                                                                        & 1.77                                                                                & 1                                                                            & 236                                                                            & 1.77                                                                        & 143                                                                          & 214                                                                            & 1.91                                                                        \\
				& 200                   & 3.09                                                                                    & 3.09                                                                        & 2.70                                                                                & 1                                                                            & 129                                                                            & 2.70                                                                        & 3                                                                            & 119                                                                            & 2.89                                                                        \\
				& 400                   & 4.17                                                                                    & 4.14                                                                        & 3.81                                                                                & 1                                                                            & 50                                                                             & 3.81                                                                        & 21                                                                           & 42                                                                             & 3.88                                                                        \\
				& 800                   & 5.70                                                                                    & 5.19                                                                        & 4.30                                                                                & 1                                                                            & 3                                                                              & 4.90                                                                        & 3                                                                            & 3                                                                              & 5.07                                                                        \\
				& 1200                  & 6.70                                                                                    & 4.80                                                                        & 4.74                                                                                & 1                                                                            & 25                                                                             & 5.20                                                                        & 18                                                                           & 23                                                                             & 5.39                                                                        \\
				& 1600                  & no                                                                                      & no                                                                          & 3.22                                                                                & 1                                                                            & 18                                                                             & 5.52                                                                        & 15                                                                           & 17                                                                             & 5.71                                                                        \\ \midrule
				\multirow{6}{*}{\begin{tabular}[c]{@{}c@{}}Ger-\\      many\end{tabular}} & 50                    & 2.06                                                                                    & 2.06                                                                        & 1.92                                                                                & 3                                                                            & 9                                                                              & 1.92                                                                        & 4                                                                            & 10                                                                             & 2.06                                                                        \\
				& 100                   & 2.63                                                                                    & 2.63                                                                        & 2.59                                                                                & 107                                                                          & 125                                                                            & 2.56                                                                        & 2                                                                            & 18                                                                             & 2.60                                                                        \\
				& 200                   & 4.21                                                                                    & 4.21                                                                        & 4.04                                                                                & 3                                                                            & 5                                                                              & 4.02                                                                        & 2                                                                            & 3                                                                              & 4.14                                                                        \\
				& 400                   & 5.50                                                                                    & 4.28                                                                        & 4.50                                                                                & 1                                                                            & 3                                                                              & 5.17                                                                        & 1                                                                            & 3                                                                              & 5.17                                                                        \\
				& 600                   & 6.60                                                                                    & 4.54                                                                        & 4.75                                                                                & 1                                                                            & 20                                                                             & 5.45                                                                        & 5                                                                            & 16                                                                             & 5.47                                                                        \\
				& 800                   & no                                                                                      & no                                                                          & 2.97                                                                                & 1                                                                            & 9                                                                              & 5.82                                                                        & 8                                                                            & 8                                                                              & 5.89                                                                        \\ \bottomrule
			\end{tabular}
			\begin{tablenotes}   
				\footnotesize               
				\item[1] Best iter: the iteration index when the best solution is found; Total iter: total number of iterations; UB sav: UB saving ratio; Sav: saving ratio; ``no" means that no feasible solution is found within the time limit.
			\end{tablenotes}
		\end{threeparttable}
		\label{table:ourHeuristic}
	\end{table}

	Table~\ref{table:ourHeuristic} illustrates the computational performance of various solution methods for TPP using a time unit of 10 minutes and a platoon size limit of 5. Two indicators, saving ratio and UB saving ratio, are employed to evaluate the performance. The saving ratio measures the cost savings achieved by using the best-found solution compared to the shortest-path solution without platooning, while the UB saving ratio represents the maximum potential cost savings that can be achieved through platooning, calculated as the difference ratio between the best-known bound and the shortest-path fuel cost without platooning.
	Since an optimal solution cannot be obtained, the performance of CPF within a 10-hour computation time frame is presented to provide a range of fuel-saving ratios. The results for TSF are not included as it is unable to produce a feasible solution for larger instances. 
	All heuristics procedures (LLHeur, LLIter, and IHeur) are halted when a recorded path in the routing stage is repeated 3 times, and when the 30-minute time limit is reached.
	Furthermore, when implementing LLIter and IHeur on large instances (Grid instance with 1200 and 1600 vehicles; German instance with 600 and 800 vehicles), we use a pairwise pre-processing heuristic (its detailed performance can be seen from Table~\ref{table:largeschedulingProblem} in the appendix) for the scheduling problem, whose parameters $\gamma$ is set to 20\%. 
	
	Overall, our implementation (IHeur) is able to obtain a near-optimal solution within 30 minutes. The results show that, compared to CPF, IHeur is highly effective. For instances with low vehicle numbers (Grid instances with 100 and 200 vehicles; German instances with 50, 100, and 200 vehicles), IHeur's solution within 30 minutes is almost optimal since its saving ratio is close to the UB saving ratio of CPF. For instances with high vehicle numbers (Grid instances with 1200 and 1600 vehicles; German instances with 400, 600, and 800 vehicles), the IHeur's solution within 30 minutes is even better than the best-found solution of CPF within 10 hours.
	
	Compared with LLHeur, IHeur is quite good across all settings. Moreover, IHeur does relatively better on the German network (with $647$ nodes) than on the grid network (with only $100$ nodes); this is also the case for \citeauthor{Luo2020}'s original implementation.  
	
	In comparing the cost updates, we isolate the effect of the two mechanisms, LLIter and IHeur, by testing them with the same routing formulation FCNF and the same scheduling formulation TIF. It is clear that IHeur (ICMP) is more effective than LLIter (LLCMP). This conclusion is further supported by analyzing the iteration index at which the best solution is found and the total number of iterations. Unlike LLIter, the best solution for IHeur is usually not found in the first iteration, highlighting IHeur's increased potential for achieving improvements across different iterations of the overall heuristic procedure.
	
	For instances without a platoon size limit, we are fortunate to obtain optimal solutions for more instances, allowing us to further demonstrate how close IHeur's solution is to optimality. Table~\ref{table:iterative} and Table~\ref{table:ourHeuristic11} in the appendix demonstrate that IHeur can achieve an optimality gap of 0.01\% to 0.05\% for German instances and 0.2\% for Grid instances. This further emphasizes the significance of the 0.1\% or 0.2\% improvement from LLIter to IHeur.
	
	Though this section cannot prove that IHeur outperforms LLHeur under all conditions, it does demonstrate that IHeur can achieve notable progress when a reasonable time unit is selected.

	\section{Conclusion}\label{conclusion}
	
	We study the truck platooning problem, in which a truck's route and schedule can be adjusted to join a platoon and thus reduce fuel consumption. First, we compare two types of MIP formulations, CPF and TSF, for TPP, where we observe that vehicle number, platoon size limit, and the coarseness of the time discretization might influence their computational performance. Then, by using the decomposition-based framework of \citet{Luo2020}, we examine different formulations for the two subproblems (routing and scheduling) and slightly improve the iterative process (the cost updates). 
	We achieve a speed-up in the computation time and further improve the platooning fuel savings under the heuristic framework. 
	A pairwise pre-processing scheduling heuristic is also implemented for forming a number of platooning pairs upfront, before solving very large scheduling instances. 
	%\textcolor{blue}{Our contributions to the scheduling problem are not only an alternative formulation (TIF) under specific scenarios, but also a pairwise heuristic framework meaningful for other formulations. Since the scheduling problem is already a realistic problem, our alternative formulation, insights, and heuristics for it are more meaningful.}
	Based on numerical experiments, we find that the resulting procedure is suitable for practical situations because it can quickly coordinate the routes and schedules for many trucks. 
	
	We design the heuristic for TPP from the perspective of third-party shared mobility platforms such as Uber or DiDi. Even though for some instances, our heuristic only accomplishes about 1\% more fuel savings than existing methods, it is effective. Since the fuel savings represents all potential revenue for the platform, an increase in fuel savings from 4\% to 5\% represents a 25\% improvement in revenue.
	\bibliographystyle{model5names}
	%\biboptions{authoryear}
	\bibliography{library2}

\begin{thebibliography}{26}
\expandafter\ifx\csname natexlab\endcsname\relax\def\natexlab#1{#1}\fi
\providecommand{\url}[1]{\texttt{#1}}
\providecommand{\href}[2]{#2}
\providecommand{\path}[1]{#1}
\providecommand{\DOIprefix}{doi:}
\providecommand{\ArXivprefix}{arXiv:}
\providecommand{\URLprefix}{URL: }
\providecommand{\Pubmedprefix}{pmid:}
\providecommand{\doi}[1]{\href{http://dx.doi.org/#1}{\path{#1}}}
\providecommand{\Pubmed}[1]{\href{pmid:#1}{\path{#1}}}
\providecommand{\bibinfo}[2]{#2}
\ifx\xfnm\relax \def\xfnm[#1]{\unskip,\space#1}\fi
%Type = Article
\bibitem[{Abdolmaleki et~al.(2021)Abdolmaleki, Shahabi, Yin \&
  Masoud}]{ABDOLMALEKI202191}
\bibinfo{author}{Abdolmaleki, M.}, \bibinfo{author}{Shahabi, M.},
  \bibinfo{author}{Yin, Y.}, \& \bibinfo{author}{Masoud, N.}
  (\bibinfo{year}{2021}).
\newblock \bibinfo{title}{{Itinerary planning for cooperative truck
  platooning}}.
\newblock {\it \bibinfo{journal}{Transportation Research Part B:
  Methodological}\/},  {\it \bibinfo{volume}{153}\/}, \bibinfo{pages}{91--110}.
  \DOIprefix\doi{10.1016/j.trb.2021.08.016}.
%Type = Techreport
\bibitem[{Albi{\'{n}}ski et~al.(2020)Albi{\'{n}}ski, Crainic \&
  Minner}]{Crainic2020}
\bibinfo{author}{Albi{\'{n}}ski, S.}, \bibinfo{author}{Crainic, T.~G.}, \&
  \bibinfo{author}{Minner, S.} (\bibinfo{year}{2020}).
\newblock {\it \bibinfo{title}{{The Day-before Truck Platooning Planning
  Problem and the Value of Autonomous Driving}}\/}.
\newblock \bibinfo{type}{Technical Report} CIRRELT.
\newblock \URLprefix
  \url{https://www.cirrelt.ca/documentstravail/cirrelt-2020-04.pdf}.
%Type = Article
\bibitem[{Bhoopalam et~al.(2020)Bhoopalam, Agatz \&
  Zuidwijk}]{KishoreBhoopalam2020}
\bibinfo{author}{Bhoopalam, A.~K.}, \bibinfo{author}{Agatz, N.~A.}, \&
  \bibinfo{author}{Zuidwijk, R.~A.} (\bibinfo{year}{2020}).
\newblock \bibinfo{title}{{Spatial and Temporal Synchronization of Truck
  Platoons}}.
\newblock {\it \bibinfo{journal}{SSRN Electronic Journal}\/},  (pp.
  \bibinfo{pages}{1--42}). \DOIprefix\doi{10.2139/ssrn.3741234}.
%Type = Inproceedings
\bibitem[{Bonnet \& Fritz(2000)}]{Bonnet2000}
\bibinfo{author}{Bonnet, C.}, \& \bibinfo{author}{Fritz, H.}
  (\bibinfo{year}{2000}).
\newblock \bibinfo{title}{{Fuel consumption reduction in a platoon:
  Experimental results with two electronically coupled trucks at close
  spacing}}.
\newblock In {\it \bibinfo{booktitle}{SAE Technical Papers}\/}.
\newblock \DOIprefix\doi{10.4271/2000-01-3056}.
%Type = Article
\bibitem[{Boysen et~al.(2018)Boysen, Briskorn \& Schwerdfeger}]{Boysen2018}
\bibinfo{author}{Boysen, N.}, \bibinfo{author}{Briskorn, D.}, \&
  \bibinfo{author}{Schwerdfeger, S.} (\bibinfo{year}{2018}).
\newblock \bibinfo{title}{{The identical-path truck platooning problem}}.
\newblock {\it \bibinfo{journal}{Transportation Research Part B:
  Methodological}\/},  {\it \bibinfo{volume}{109}\/}, \bibinfo{pages}{26--39}.
  \DOIprefix\doi{10.1016/j.trb.2018.01.006}.
%Type = Article
\bibitem[{Christofides et~al.(1987)Christofides, Alvarez-Valdes \&
  Tamarit}]{Christofides1987}
\bibinfo{author}{Christofides, N.}, \bibinfo{author}{Alvarez-Valdes, R.}, \&
  \bibinfo{author}{Tamarit, J.~M.} (\bibinfo{year}{1987}).
\newblock \bibinfo{title}{{Project scheduling with resource constraints: A
  branch and bound approach}}.
\newblock {\it \bibinfo{journal}{European Journal of Operational Research}\/},
  {\it \bibinfo{volume}{29}\/}, \bibinfo{pages}{262--273}.
  \DOIprefix\doi{10.1016/0377-2217(87)90240-2}.
%Type = Article
\bibitem[{Creemers et~al.(2017)Creemers, Woumans, Boute \&
  Beli{\"{e}}n}]{Creemers2017}
\bibinfo{author}{Creemers, S.}, \bibinfo{author}{Woumans, G.},
  \bibinfo{author}{Boute, R.}, \& \bibinfo{author}{Beli{\"{e}}n, J.}
  (\bibinfo{year}{2017}).
\newblock \bibinfo{title}{{Tri-vizor uses an efficient algorithm to identify
  collaborative shipping opportunities}}.
\newblock {\it \bibinfo{journal}{Interfaces}\/},  {\it \bibinfo{volume}{47}\/},
  \bibinfo{pages}{244--259}. \DOIprefix\doi{10.1287/inte.2016.0878}.
%Type = Article
\bibitem[{Dahle et~al.(2019)Dahle, Andersson, Christiansen \&
  Speranza}]{Dahle2019}
\bibinfo{author}{Dahle, L.}, \bibinfo{author}{Andersson, H.},
  \bibinfo{author}{Christiansen, M.}, \& \bibinfo{author}{Speranza, M.~G.}
  (\bibinfo{year}{2019}).
\newblock \bibinfo{title}{{The pickup and delivery problem with time windows
  and occasional drivers}}.
\newblock {\it \bibinfo{journal}{Computers and Operations Research}\/},  {\it
  \bibinfo{volume}{109}\/}, \bibinfo{pages}{122--133}.
  \DOIprefix\doi{10.1016/j.cor.2019.04.023}.
%Type = Misc
\bibitem[{{European Automobile Manufacturers'
  Association}(2017)}]{TheEuropeanAutomobileManufacturersAssociationACEA2017}
\bibinfo{author}{{European Automobile Manufacturers' Association}}
  (\bibinfo{year}{2017}).
\newblock \bibinfo{title}{{What Is Truck Platooning?}}
\newblock \URLprefix
  \url{https://www.acea.auto/files/Platooning{\_}roadmap.pdf}.
%Type = Inproceedings
\bibitem[{Janssen et~al.(2015)Janssen, Zwijnenberg, Blankers \&
  de~Kruijff}]{Janssen2015}
\bibinfo{author}{Janssen, G.~R.}, \bibinfo{author}{Zwijnenberg, H.},
  \bibinfo{author}{Blankers, I.}, \& \bibinfo{author}{de~Kruijff, J.}
  (\bibinfo{year}{2015}).
\newblock \bibinfo{title}{{Truck Platooning: driving the Future of
  Transportation}}.
\newblock In {\it \bibinfo{booktitle}{TNO whitepaper}\/}.
\newblock \URLprefix
  \url{http://resolver.tudelft.nl/uuid:778397eb-59d3-4d23-9185-511385b91509}.
%Type = Article
\bibitem[{Larsen et~al.(2019)Larsen, Rich \& Rasmussen}]{Larsen2019}
\bibinfo{author}{Larsen, R.}, \bibinfo{author}{Rich, J.}, \&
  \bibinfo{author}{Rasmussen, T.~K.} (\bibinfo{year}{2019}).
\newblock \bibinfo{title}{{Hub-based truck platooning: Potentials and
  profitability}}.
\newblock {\it \bibinfo{journal}{Transportation Research Part E: Logistics and
  Transportation Review}\/},  {\it \bibinfo{volume}{127}\/},
  \bibinfo{pages}{249--264}. \DOIprefix\doi{10.1016/j.tre.2019.05.005}.
%Type = Inproceedings
\bibitem[{Larson et~al.(2013)Larson, Kammer, Liang \& Johansson}]{Larson2013}
\bibinfo{author}{Larson, J.}, \bibinfo{author}{Kammer, C.},
  \bibinfo{author}{Liang, K.~Y.}, \& \bibinfo{author}{Johansson, K.~H.}
  (\bibinfo{year}{2013}).
\newblock \bibinfo{title}{{Coordinated route optimization for heavy-duty
  vehicle platoons}}.
\newblock In {\it \bibinfo{booktitle}{IEEE Conference on Intelligent
  Transportation Systems, Proceedings}\/} (pp. \bibinfo{pages}{1196--1202}).
\newblock \DOIprefix\doi{10.1109/ITSC.2013.6728395}.
%Type = Inproceedings
\bibitem[{Larson et~al.(2016)Larson, Munson \& Sokolov}]{Larson2016}
\bibinfo{author}{Larson, J.}, \bibinfo{author}{Munson, T.}, \&
  \bibinfo{author}{Sokolov, V.} (\bibinfo{year}{2016}).
\newblock \bibinfo{title}{{Coordinated Platoon Routing in a Metropolitan
  Network}}.
\newblock In {\it \bibinfo{booktitle}{2016 Proceedings of the Seventh SIAM
  Workshop on Combinatorial Scientific Computing}\/} (pp.
  \bibinfo{pages}{73--82}).
\newblock \DOIprefix\doi{10.1137/1.9781611974690.ch8}.
%Type = Article
\bibitem[{Larsson et~al.(2015)Larsson, Sennton \& Larson}]{Larsson2015}
\bibinfo{author}{Larsson, E.}, \bibinfo{author}{Sennton, G.}, \&
  \bibinfo{author}{Larson, J.} (\bibinfo{year}{2015}).
\newblock \bibinfo{title}{{The vehicle platooning problem: Computational
  complexity and heuristics}}.
\newblock {\it \bibinfo{journal}{Transportation Research Part C: Emerging
  Technologies}\/},  {\it \bibinfo{volume}{60}\/}, \bibinfo{pages}{258--277}.
  \DOIprefix\doi{10.1016/j.trc.2015.08.019}.
%Type = Article
\bibitem[{Liang et~al.(2016)Liang, M{\aa}rtensson \& Johansson}]{Liang2016}
\bibinfo{author}{Liang, K.~Y.}, \bibinfo{author}{M{\aa}rtensson, J.}, \&
  \bibinfo{author}{Johansson, K.~H.} (\bibinfo{year}{2016}).
\newblock \bibinfo{title}{{Heavy-Duty Vehicle Platoon Formation for Fuel
  Efficiency}}.
\newblock {\it \bibinfo{journal}{IEEE Transactions on Intelligent
  Transportation Systems}\/},  {\it \bibinfo{volume}{17}\/},
  \bibinfo{pages}{1051--1061}. \DOIprefix\doi{10.1109/TITS.2015.2492243}.
%Type = Article
\bibitem[{Luo(2022)}]{Luo2022}
\bibinfo{author}{Luo, F.} (\bibinfo{year}{2022}).
\newblock \bibinfo{title}{{Coordinated Vehicle Platooning with Fixed Routes:
  Adaptive Time Discretization, Strengthened Formulations and Approximation
  Algorithms}}.
\newblock {\it \bibinfo{journal}{SSRN Electronic Journal}\/},  (pp.
  \bibinfo{pages}{1--43}). \DOIprefix\doi{10.2139/ssrn.4116041}.
%Type = Article
\bibitem[{Luo \& Larson(2022)}]{Luo2020}
\bibinfo{author}{Luo, F.}, \& \bibinfo{author}{Larson, J.}
  (\bibinfo{year}{2022}).
\newblock \bibinfo{title}{{A Repeated Route-then-Schedule Approach to
  Coordinated Vehicle Platooning: Algorithms, Valid Inequalities and
  Computation}}.
\newblock {\it \bibinfo{journal}{Operations Research}\/},  {\it
  \bibinfo{volume}{70}\/}, \bibinfo{pages}{2477--2495}.
  \DOIprefix\doi{10.1287/opre.2021.2126}.
%Type = Article
\bibitem[{Luo et~al.(2018)Luo, Larson \& Munson}]{Luo2018}
\bibinfo{author}{Luo, F.}, \bibinfo{author}{Larson, J.}, \&
  \bibinfo{author}{Munson, T.} (\bibinfo{year}{2018}).
\newblock \bibinfo{title}{{Coordinated platooning with multiple speeds}}.
\newblock {\it \bibinfo{journal}{Transportation Research Part C: Emerging
  Technologies}\/},  {\it \bibinfo{volume}{90}\/}, \bibinfo{pages}{213--225}.
  \DOIprefix\doi{10.1016/j.trc.2018.02.011}.
%Type = Misc
\bibitem[{{Ministry of Transport - Singapore}(2017)}]{Tekst2017}
\bibinfo{author}{{Ministry of Transport - Singapore}} (\bibinfo{year}{2017}).
\newblock \bibinfo{title}{{Singapore to Start Truck Platooning Trials}}.
\newblock \URLprefix \url{https//www.mot.gov.sg/news-centre/news}.
%Type = Article
\bibitem[{Miranda et~al.(2022)Miranda, Cordeau \& Frejinger}]{Miranda2022}
\bibinfo{author}{Miranda, P.~L.}, \bibinfo{author}{Cordeau, J.~F.}, \&
  \bibinfo{author}{Frejinger, E.} (\bibinfo{year}{2022}).
\newblock \bibinfo{title}{{A time–space formulation for the locomotive
  routing problem at the Canadian National Railways}}.
\newblock {\it \bibinfo{journal}{Computers and Operations Research}\/},  {\it
  \bibinfo{volume}{139}\/}, \bibinfo{pages}{105629}.
  \DOIprefix\doi{10.1016/j.cor.2021.105629}.
%Type = Inproceedings
\bibitem[{Nourmohammadzadeh \& Hartmann(2016)}]{10.1007/978-3-319-49001-4_4}
\bibinfo{author}{Nourmohammadzadeh, A.}, \& \bibinfo{author}{Hartmann, S.}
  (\bibinfo{year}{2016}).
\newblock \bibinfo{title}{{The Fuel-Efficient Platooning of Heavy Duty Vehicles
  by Mathematical Programming and Genetic Algorithm}}.
\newblock In {\it \bibinfo{booktitle}{International Conference on Theory and
  Practice of Natural Computing}\/} (pp. \bibinfo{pages}{46--57}).
\newblock \DOIprefix\doi{10.1007/978-3-319-49001-4_4}.
%Type = Article
\bibitem[{Nourmohammadzadeh \& Hartmann(2019)}]{Nourmohammadzadeh2019}
\bibinfo{author}{Nourmohammadzadeh, A.}, \& \bibinfo{author}{Hartmann, S.}
  (\bibinfo{year}{2019}).
\newblock \bibinfo{title}{{Fuel-efficient truck platooning by a novel
  meta-heuristic inspired from ant colony optimisation}}.
\newblock {\it \bibinfo{journal}{Soft Computing}\/},  {\it
  \bibinfo{volume}{23}\/}, \bibinfo{pages}{1439--1452}.
  \DOIprefix\doi{10.1007/s00500-018-3518-x}.
%Type = Techreport
\bibitem[{Tiernan et~al.(2017)Tiernan, Richardson, Azeredo, Najm \&
  Lochrane}]{Tiernan2017}
\bibinfo{author}{Tiernan, T.~A.}, \bibinfo{author}{Richardson, N.},
  \bibinfo{author}{Azeredo, P.}, \bibinfo{author}{Najm, W.~G.}, \&
  \bibinfo{author}{Lochrane, T.} (\bibinfo{year}{2017}).
\newblock {\it \bibinfo{title}{{Test and Evaluation of Vehicle Platooning
  Proof-of-Concept Based on Cooperative Adaptive Cruise Control Final
  Report}}\/}.
\newblock \bibinfo{type}{Technical Report} US Department of Transportation.
\newblock \URLprefix \url{https://rosap.ntl.bts.gov/view/dot/1038}.
%Type = Phdthesis
\bibitem[{{Van de Hoef}(2016)}]{VandeHoef2016}
\bibinfo{author}{{Van de Hoef}, S.} (\bibinfo{year}{2016}).
\newblock {\it \bibinfo{title}{{Fuel-Efficient Centralized Coordination of
  Truck Platooning}}\/}.
\newblock \bibinfo{type}{Licentiate thesis.} KTH Royal Institute of Technology.
\newblock \URLprefix
  \url{https://people.kth.se/~kallej/grad_students/vdhoef_licthesis16.pdf}.
%Type = Article
\bibitem[{Zhang et~al.(2017)Zhang, Jenelius \& Ma}]{Zhang2017}
\bibinfo{author}{Zhang, W.}, \bibinfo{author}{Jenelius, E.}, \&
  \bibinfo{author}{Ma, X.} (\bibinfo{year}{2017}).
\newblock \bibinfo{title}{{Freight transport platoon coordination and departure
  time scheduling under travel time uncertainty}}.
\newblock {\it \bibinfo{journal}{Transportation Research Part E: Logistics and
  Transportation Review}\/},  {\it \bibinfo{volume}{98}\/},
  \bibinfo{pages}{1--23}. \DOIprefix\doi{10.1016/j.tre.2016.11.008}.
%Type = Article
\bibitem[{Zhen et~al.(2019)Zhen, Wang, Laporte \& Hu}]{Zhen2019}
\bibinfo{author}{Zhen, L.}, \bibinfo{author}{Wang, S.},
  \bibinfo{author}{Laporte, G.}, \& \bibinfo{author}{Hu, Y.}
  (\bibinfo{year}{2019}).
\newblock \bibinfo{title}{{Integrated planning of ship deployment, service
  schedule and container routing}}.
\newblock {\it \bibinfo{journal}{Computers and Operations Research}\/},  {\it
  \bibinfo{volume}{104}\/}, \bibinfo{pages}{304--318}.
  \DOIprefix\doi{10.1016/j.cor.2018.12.022}.

\end{thebibliography}
	\setcounter{table}{0}
	
	\newpage
	
	\appendix
	\section{Abbreviations}
	
	\numberwithin{table}{section}
	\begin{table}[H]
		\footnotesize   
		\centering
		\caption{Summary of abbreviations}
		\begin{tabular}{@{}lll@{}}
			\toprule
			Abbreviation & Meaning                                                                                                                                                            &  \\ \midrule
			TPP          & Truck platooning problem                                                                                                                                               &  \\
			LRF          & The routing formulation of \citet{Luo2020}                                                                                                            &  \\
			FCNF         & 
			Our routing formulation (fixed-charge network flow problem)
			\\
			ABF          & The scheduling formulation of \citet{Luo2020},   assignment-based formulation                                                                         &  \\
			TIF          & Our scheduling formulation, time-indexed formulation                                                                                                                   &  \\
			LLCMP        & The cost modification procedure of \citet{Luo2020}                                                                                                    &  \\
			ICMP         & Our improved cost modification procedure  &  \\ 
			LLIter        & \citet{Luo2020}'s iteration, FCNF + TIF + LLCMP  &  \\
			LLHeur         & \citet{Luo2020}'s heuristic, LRF + ABF + LLCMP &  \\ 
			IHeur         & Our improved heuristic, FCNF + TIF + ICMP&  \\ 
			\bottomrule
		\end{tabular}\label{table:fnotation}
	\end{table}

	\begin{table}[H]
		\footnotesize   
		\centering
		\caption{Summary of indicators}
		\begin{tabular}{@{}lll@{}}
			\toprule
			Indicators & Explanation                                                                                                                                                            &  \\ \midrule                                                                   
			Optimality gap         &$\left|Obj-Opt\right|/Opt$, the percentage difference between the best-found solution \\&value  ($Obj$) and the optimal objective value ($Opt$) &  \\ 
			Relative gap        &$\left|Obj-Bound\right|/Bound$, the percentage difference between the best-found solution \\&value  and the best-known bound ($Bound$)  within the time limit &  \\
			Saving ratio         & $\left|SPC-Obj\right|/SPC$, the ratio of the cost savings achieved by using the best-found  \\&solution to the shortest-path fuel cost without platooning ($SPC$) &  \\ 
			UB saving ratio         & $\left|SPC-Bound\right|/SPC$, the maximum potential cost savings, the difference ratio  \\&between the best-known bound and the shortest-path fuel cost  without platooning &  \\ 
			\bottomrule
		\end{tabular}\label{table:indicatornotation}
	\end{table}

	\section{Computational performance of pairwise scheduling heuristic}
	\begin{table}[h]
		\footnotesize   
		\centering
		\caption{Computational performance for large scheduling problem with platoon size $Q=5$ and time unit $TU=10$ min} \label{table:resultsschedulinglarge}
		\begin{threeparttable} 
			\begin{tabular}{@{}crrrrrr@{}}
				\toprule
				& \multicolumn{1}{c}{}        & \multicolumn{1}{c}{}            & \multicolumn{2}{c}{ABF}                                    & \multicolumn{2}{c}{TIF (heuristic)}                        \\ \cmidrule(l){4-7} 
				Network                 & \multicolumn{1}{c}{Vehicle} & \multicolumn{1}{c}{UB Sav (\%)} & \multicolumn{1}{c}{Sav (\%)} & \multicolumn{1}{c}{CPU (s)} & \multicolumn{1}{c}{Sav (\%)} & \multicolumn{1}{c}{CPU (s)} \\ \midrule
				\multirow{5}{*}{Grid}   & 800                         & 5.34                            & 3.85                         & 1800.0                      & 4.63                         & 6.5                         \\
				& 1000                        & 5.76                            & 4.62                         & 1800.0                      & 4.88                         & 8.3                         \\
				& 1200                        & 6.26                            & 4.86                         & 1800.0                      & 5.19                         & 12.2                        \\
				& 1400                        & 6.60                            & 4.96                         & 1800.0                      & 5.40                         & 15.3                        \\
				& 1600                        & 6.84                            & 5.10                         & 1800.0                      & 5.55                         & 21.8                        \\ \midrule
				\multirow{5}{*}{German} & 400                         & 5.54                            & 0.00                         & 1800.0                      & 4.84                         & 12.2                        \\
				& 500                         & 5.89                            & 4.56                         & 1800.0                      & 5.14                         & 16.9                        \\
				& 600                         & 6.53                            & 4.85                         & 1800.0                      & 5.46                         & 33.8                        \\
				& 700                         & 6.77                            & 2.65                         & 1800.0                      & 5.55                         & 63.3                        \\
				& 800                         & 7.19                            & 2.96                         & 1800.0                      & 5.82                         & 104.6                       \\ \bottomrule
			\end{tabular}	
		\end{threeparttable} 
		\label{table:largeschedulingProblem}
	\end{table}
	
	Table~\ref{table:largeschedulingProblem} contains the results for large instances with the platoon size $Q=5$ and time unit $TU=10$ min. The large instances are the Grid instances with at least 800 vehicles and German instances with at least 400 vehicles. This table aims to illustrate the effectiveness of pairwise scheduling heuristics for large instances (for medium-size instances, we only use the solver to get results). Here, ABF is solved by the solver, while TIF is run with the pairwise pre-processing heuristic, where the pre-processing parameter $\gamma$ is set to 20\%. 
	Two indicators are used: UB saving ratio and saving ratio. UB saving ratio is the percentage difference between the ABF's best-known bound within 2 hours and the shortest path fuel cost. The saving ratio is the percentage difference between the best-found solution's objective value within 30 minutes and the shortest path fuel cost.
	The table shows that TIF in combination with the pre-processing heuristic yields significantly better results, both in terms of solution quality as well as computation time. 
	
	\section{Computational performance for scheduling problem and TPP without platoon size limit} 
	
	%Several indicators are used in this section, including the optimality gap, the relative gap, and the saving ratio.	The relative gap is computed as the percentage difference between the best-found solution and the best-known bound on the solution objective within the time limit.  The optimality gap is the percentage by which the objective value of the best-found solution exceeds the optimal objective value (the latter is obtained by running a formulation to completion, regardless of the runtime). The saving ratio measures the cost savings achieved by using the best-found solution compared to the shortest-path solution without platooning.
	
	\subsection{Performance of different formulations for scheduling problem without platoon size limit} \label{subsub:scheduling22}
	
	\begin{table}[H]
		\footnotesize   
		\centering
		\caption{Computational performance for medium-size instances of scheduling problem without platoon size limit}\label{schedulingResult22}
		\begin{threeparttable}
			\begin{tabular}{@{}cccccccc@{}}
				\toprule
				\multicolumn{1}{l}{}                                                          & \multicolumn{1}{l}{} & \multicolumn{3}{c}{ABF}                                                                 & \multicolumn{3}{c}{TIF}                                                                                                       \\\cmidrule(l){3-8} 
				Network                                                                       & Vehicles                  & \begin{tabular}[c]{@{}c@{}}Relative\\ gap (\%)%\tnote{1}
					\\      (10 min)\end{tabular} & \begin{tabular}[c]{@{}c@{}}Relative\\ gap (\%)\\      (30 min)\end{tabular} & CPU (s) & \begin{tabular}[c]{@{}c@{}}Relative\\ gap (\%)\\      (10 min)\end{tabular} & \begin{tabular}[c]{@{}c@{}}Relative \\gap (\%)\\      (30 min)\end{tabular} & CPU (s) \\ \midrule
				\multirow{5}{*}{\begin{tabular}[c]{@{}c@{}}Grid     \end{tabular}}   & 200                  & 0.00                                                         & 0.00                                                         & 0.5     & 0.00                                                         & 0.00                                                         & 0.2     \\
				& 400                  & 0.00                                                         & 0.00                                                         & 88.7    & 0.00                                                         & 0.00                                                         & 1.3     \\
				& 600                  & 0.60                                                         & 0.36                                                         & 1800.0  & 0.00                                                         & 0.00                                                         & 13.8    \\
				& 800                  & 3.67                                                         & 2.07                                                         & 1800.0  & 0.00                                                         & 0.00                                                         & 189.9   \\
				& 1000                 & 18.94                                                        & 17.88                                                        & 1800.0  & 1.59                                                         & 0.85                                                         & 1800.0    \\\cmidrule(l){1-8} 
				\multirow{5}{*}{\begin{tabular}[c]{@{}c@{}}German\end{tabular}} & 100                  & 0.00                                                         & 0.00                                                         & 0.1     & 0.00                                                         & 0.00                                                         & 0.3     \\
				& 200                  & 0.00                                                         & 0.00                                                         & 4.8     & 0.00                                                         & 0.00                                                         & 4.1     \\
				& 300                  & 0.24                                                         & 0.02                                                         & 1800.0  & 0.00                                                         & 0.00                                                         & 13.2    \\
				& 400                  & 14.03                                                        & 1.55                                                         & 1800.0  & 0.00                                                         & 0.00                                                         & 269.6   \\
				& 500                  & 17.27                                                        & 16.07                                                        & 1800.0  & no\tnote{1}                                                            & 0.00                                                         & 757.4   \\ 
				\bottomrule
			\end{tabular}
			\begin{tablenotes}   
				\footnotesize     
				\item[1] no feasible solution       
			\end{tablenotes}
		\end{threeparttable} 
		\label{table:schedulingProblem22}
	\end{table}
	
	Table~\ref{table:schedulingProblem22} contains the results of the scheduling problem without the platoon size limit.	The lower relative gap and CPU time illustrate the TIF performs better than ABF in time and solution quality.

	\subsection{Performance of different choices for the cost updates without considering platoon size limit} \label{subsub:compare-iterative}
	
	\begin{table}[H]
		\footnotesize 
		\centering
		\caption{Comparison of the cost updates via LLIter and IHeur}
		\begin{threeparttable}
			\begin{tabular}{@{}cccccccccc@{}}
				\toprule
				\multicolumn{1}{l}{}                                                     & \multicolumn{1}{l}{}    & \multicolumn{4}{c}{LLIter (LLCMP)}                                                                                                                                        & \multicolumn{4}{c}{IHeur (ICMP)}                                                                                                            \\ \cmidrule(l){3-10}
				\multicolumn{1}{l}{Network}                                              & \multicolumn{1}{l}{Vehicle} & \begin{tabular}[c]{@{}c@{}}CPU\\ time (s) \end{tabular} & \begin{tabular}[c]{@{}c@{}}Optimality\\ gap (\%)\\ (30 min)\end{tabular} & \multicolumn{1}{l}{\begin{tabular}[c]{@{}c@{}}Best\\ iter\tnote{1} \end{tabular}} & \multicolumn{1}{l}{\begin{tabular}[c]{@{}c@{}}Total\\ iter\tnote{2} \end{tabular}} & \begin{tabular}[c]{@{}c@{}}CPU\\ time (s) \end{tabular} & \begin{tabular}[c]{@{}c@{}}Optimality\\ gap (\%)\\ (30 min)\end{tabular} & \begin{tabular}[c]{@{}c@{}}Best\\ iter \end{tabular} & \begin{tabular}[c]{@{}c@{}}Total\\ iter \end{tabular}  \\\midrule
				\multirow{5}{*}{\begin{tabular}[c]{@{}c@{}}Grid\end{tabular}}   & 200                     & 1800.0                                                                              & 0.41                                                                  & 1                             & 115                            & 1800.0                                                                                  & 0.22                                                                                      & 3                             & 108                            \\
				& 400                     & 1800.0                                                                              & 0.35                                                                  & 1                             & 67                             & 1800.0                                                                                  & 0.24                                                                                      & 29                            & 60                             \\
				& 600                     & 1800.0                                                                              & 0.42                                                                  & 1                             & 46                             & 1800.0                                                                                  & 0.24                                                                                      & 39                            & 42                             \\
				& 800                     & 1800.0                                                                              & 0.46                                                                  & 1                             & 7                              & 1800.0                                                                                  & 0.26                                                                                      & 7                             & 11                             \\
				& 1000                    & 1800.0                                                                              & 0.45                                                                  & 1                             & 3                              & 1800.0                                                                                  & 0.29                                                                                      & 3                             & 3                              \\\midrule
				\multirow{5}{*}{\begin{tabular}[c]{@{}c@{}}German\end{tabular}} & 100                     & 482.4                                                                               & 0.05                                                                  & 54                            & 57                             & 179.4                                                                                   & 0.03                                                                                      & 2                             & 28                             \\
				& 200                     & 1800.0                                                                              & 0.13                                                                  & 38                            & 57                             & 1269.7                                                                                  & 0.01                                                                                      & 16                            & 43                             \\
				& 300                     & 1800.0                                                                              & 0.18                                                                  & 1                             & 32                             & 1800.0                                                                                  & 0.05                                                                                      & 12                            & 29                             \\
				& 400                     & 1800.0                                                                              & 0.09                                                                  & 1                             & 7                              & 1800.0                                                                                  & 0.04                                                                                      & 3                             & 6                              \\
				& 500                     & 1800.0                                                                              & 0.08                                                                  & 1                             & 2                              & 1800.0                                                                                  & 0.05                                                                                     & 2                             & 2                             
				\\\bottomrule 
			\end{tabular}
			\begin{tablenotes}   
				\footnotesize               
				\item[1] Iteration index when the best solution is found.  
				\item[2] Total number of iterations.
			\end{tablenotes}
		\end{threeparttable} 
		\label{table:iterative}
	\end{table}
	The indicators include the total iteration index, the best iteration index, and the optimality gap.
	
	Table~\ref{table:iterative} compares the cost updating mechanisms LLIter and IHeur\@. In order to isolate the effect of these two mechanisms, we test the procedure with the same routing formulation FCNF and the same scheduling formulation TIF\@. 
	Both procedures are halted when a recorded path in the routing stage is repeated 3 times, and when the 30-minute time limit is reached.
	We find that IHeur requires less computation time and leads to a lower optimality gap than LLIter, so this clearly constitutes an improvement.
	This observation is reinforced by examining the iteration index when the best solution is found and the total number of iterations.  Contrary to LLIter, the best solution of IHeur is usually not found in the first iteration, which underlines the increased potential of IHeur over LLIter of achieving improvements across the different iterations of the overall heuristic procedure.

	\subsection{Performance of different solution methods for TPP without platoon size limit}\label{subsub:heuristics2}

	Table~\ref{table:ourHeuristic11} and Table~\ref{table:largeplatoon} illustrate the computational performance of TPP for medium-size instances and large-size instances without considering platoon size limit respectively. The former is measured by the optimality gap, while the latter is indicated by the saving ratio. Both two tables illustrate the advantage of our implementation (IHeur) since the optimal gap of IHeur is lower than others in Table~\ref{table:ourHeuristic11} and the saving ratio of IHeur is higher than others in Table~\ref{table:largeplatoon}.
	
	\begin{table}[H]
		\footnotesize 
		\centering
		\caption{Optimality gap (\%) of different solution methods for medium-sized instances of TPP}
		\begin{threeparttable}
			\begin{tabular}{@{}cccccccc@{}}
				\toprule
				&      & \begin{tabular}[c]{@{}c@{}}CPF\end{tabular} & \begin{tabular}[c]{@{}c@{}}TSF\end{tabular}      & LLHeur & \multicolumn{3}{c}{IHeur}                                                                                                                                                                                       \\\cmidrule(l){3-8}
				Network                                                                       & Vehicles  & \begin{tabular}[c]{@{}c@{}}30 min\end{tabular}  & \begin{tabular}[c]{@{}c@{}}30 min\end{tabular} & \begin{tabular}[c]{@{}c@{}}30 min\end{tabular}      & \begin{tabular}[c]{@{}c@{}}5 min\end{tabular} & \begin{tabular}[c]{@{}c@{}}10 min\end{tabular} & \begin{tabular}[c]{@{}c@{}}30 min\end{tabular} \\\midrule
				\multirow{5}{*}{\begin{tabular}[c]{@{}c@{}}Grid\end{tabular}}   & 200  & 0.00                                                                 & 0.00                                                                 & 0.41                                                                                          & 0.22                                                                & 0.22                                                                 & 0.22                                                                 \\
				& 400  & 0.01                                                                 & 0.00                                                                 & 0.35                                                                                          & 0.27                                                                & 0.24                                                                 & 0.24                                                                 \\
				& 600  & 0.11                                                                 & \textless{}0.01                                                      & 0.42                                                                                          & 0.27                                                                & 0.27                                                                 & 0.24                                                                 \\
				& 800  & 0.51                                                                 & no (0.55 h)\tnote{1}                                                          & 0.52                                                                                          & 0.34                                                                & 0.30                                                                 & 0.26                                                                 \\
				& 1000 & 0.53                                                                 & no (1.93 h)                                                            & 1.15                                                                                          & no                                                                    & 0.45                                                                 & 0.29                                                                 \\\midrule
				\multirow{5}{*}{\begin{tabular}[c]{@{}c@{}}German\end{tabular}} & 100  & 0.00                                                                 & 0.00                                                                 & 0.06                                                                                          & 0.03                                                                & 0.03                                                                  & 0.03                                                                 \\
				& 200  & 0.00                                                                 & 0.00                                                                 & 0.17                                                                                          & 0.01                                                                & 0.01                                                                 & 0.01                                                                 \\
				& 300  & 0.06                                                                 & 0.00                                                                 & 0.17                                                                                          & 0.07                                                                & 0.05                                                                 & 0.05                                                                 \\
				& 400  & 0.14                                                                 & no (4.94h)                                                             & 0.60                                                                                          & 0.09                                                                & 0.08                                                                 & 0.04                                                                 \\
				& 500  & 0.88                                                                 & no (15.59h)                                                            & 0.86                                                                                          & no                                                                    & no                                                                     & 0.05                                                                 \\  \bottomrule 
			\end{tabular}
			\begin{tablenotes}   
				\footnotesize               
				\item[1] ``no" means that no feasible solution is found within the time limit; between parentheses is the the time when the first feasible solution is encountered.
			\end{tablenotes}
		\end{threeparttable}
		\label{table:ourHeuristic11}
	\end{table}
	
	\begin{table}[H]
		\footnotesize   
		\centering
		\caption{Saving ratio (\%) of different solution methods for large instances of TPP}
		\begin{threeparttable} 
			\begin{tabular}{@{}cccccccc@{}}
				\toprule
				\multicolumn{1}{l}{}                                                     & \multicolumn{1}{l}{}    & \multicolumn{2}{c}{CPF}                         & LLHeur & \multicolumn{3}{c}{IHeur}   \\ \cmidrule(l){3-8}
				\multicolumn{1}{l}{Network}                                              & \multicolumn{1}{l}{Vehicle} & \begin{tabular}[c]{@{}c@{}}Total\\ time (h)\end{tabular} & \begin{tabular}[c]{@{}c@{}}30 min\end{tabular} & \begin{tabular}[c]{@{}c@{}}30 min\end{tabular} & \begin{tabular}[c]{@{}c@{}}5 min\end{tabular} & \begin{tabular}[c]{@{}c@{}}10 min\end{tabular} & \begin{tabular}[c]{@{}c@{}}30 min\end{tabular} \\\midrule
				\multirow{6}{*}{\begin{tabular}[c]{@{}c@{}}Grid\end{tabular}}   
				& 1000 & 0.59           & 5.27                                                       & 4.69                                                       & 5.25                                                      & 5.25                                                       & 5.26                                                       \\
				& 1200 & 0.66           & 5.58                                                       & 4.92                                                       & 5.65                                                      & 5.65                                                       & 5.68                                                       \\
				& 1400 & 0.88           & 5.79                                                       & 5.19                                                       & 5.81                                                      & 5.83                                                       & 5.84                                                       \\
				& 1600 & 0.82           & 6.01                                                       & 5.93                                                       & 5.98                                                      & 6.03                                                       & 6.03                                                       \\
				& 1800 & 1.00           & 5.52                                                       & 5.72                                                       & 6.07                                                      & 6.22                                                       & 6.25                                                       \\
				& 2000 & 1.18           & $-4.13$                                                      & 5.76                                                       & no\tnote{1}                                                          & no                                                           & 6.36                                                       \\\midrule
				\multirow{6}{*}{\begin{tabular}[c]{@{}c@{}}German\end{tabular}} & 500  & 0.82           & 4.82                                                       & 4.78                                                       & 5.40                                                      & 5.40                                                       & 5.40                                                       \\
				& 600  & 1.00           & 5.07                                                       & 5.17                                                       & 5.66                                                      & 5.68                                                       & 5.68                                                       \\
				& 700  & 1.39           & 5.10                                                       & 5.18                                                       & 5.75                                                      & 5.82                                                       & 5.85                                                       \\
				& 800  & 2.23           & 5.66                                                           & 5.45                                                       & 6.16                                                      & 6.16                                                       & 6.22                                                       \\
				& 900  & no           & no                                                           & 5.69                                                       & no                                                          & 6.27                                                       & 6.32                                                       \\
				& 1000 & no           & no                                                           & 5.76                                                       & no                                                          & no                                                           & 6.46                                                      
				\\ \bottomrule 
			\end{tabular}
			\begin{tablenotes}   
				\footnotesize   
				\item[1] ``no" means that no feasible solution is found within the time limit. 
			\end{tablenotes}
		\end{threeparttable} 
		\label{table:largeplatoon}
	\end{table}
	
\end{document}